\documentclass[11pt,letterpaper,fleqn]{article}
\usepackage{graphicx}
\usepackage{tabulary}
\usepackage{lineno}
\usepackage{hyperref}
\usepackage{subcaption}
\usepackage{cite}
\usepackage{tcolorbox}
\usepackage{xcolor,colortbl}
\usepackage{wrapfig}
\usepackage{array}
\usepackage{enumitem}
\usepackage{tikz}
\usetikzlibrary{positioning,calc}

\selectcolormodel{rgb}
\definecolor{mslightblue}{HTML}{DFEAF7}
\definecolor{msdarkblue}{HTML}{1D2F5A}
\definecolor{verylightgray}{rgb}{0.88, 0.88, 0.88}
\definecolor{darkgray}{rgb}{0.23, 0.23, 0.23}

\let\oldenumerate\itemize
\renewcommand{\itemize}{
  \oldenumerate
  \setlength{\itemsep}{1pt}
  \setlength{\parskip}{1pt}
  \setlength{\parsep}{1pt}
}

\newlength{\bibitemsep}
\newlength{\bibparskip}
\let\oldthebibliography\thebibliography
\renewcommand\thebibliography[1]{%
  \oldthebibliography{#1}%
  \setlength{\parskip}{\bibitemsep}%
  \setlength{\itemsep}{\bibparskip}%
}

\def\s2i2{$S^2 I^2$}

\newcommand{\statusnoteoff}[1]{}

\newcommand{\docversion}{1.00}

\title{{\bf Building an AI-native Research Ecosystem \\ for Experimental Particle Physics:  \\ A Community Vision}}
\date{\today~- Version \docversion}

\usepackage{authblk}
\author{Thea Klaeboe Aarrestad\\ETH Zürich}
\author{Alaa Abdelhamid\\University of Tennessee, Knoxville}
\author{Haider Abidi\\Brookhaven National Laboratory}
\author{Jahred Adelman\\Northern Illinois University}
\author{Jennifer Adelman-McCarthy\\Fermi National Accelerator Laboratory}
\author{Shuchin Aeron\\Tufts University}
\author{Garvita Agarwal\\University of Notre Dame}
\author{Usman Ali\\University of Oklahoma}
\author{Cristiano Alpigiani\\University of Washington}
\author{Omar Alterkait\\Tufts University}
% ...

\usepackage{datatool}

\newcommand{\roleEditor}{$^{\boldsymbol{\ast}\!\boldsymbol{\ast}}$}
\newcommand{\roleWriter}{$^{\boldsymbol{\ast}}$}
\begin{document}
%\linenumbers
\begingroup
\pagestyle{empty}
\vbox{
    \centering
    \includegraphics[width=1.0\textwidth, alt={LHCb, EIC, ATLAS, NOvA, DUNE, CMS, MicroBooNE, IceCube, XENON}]{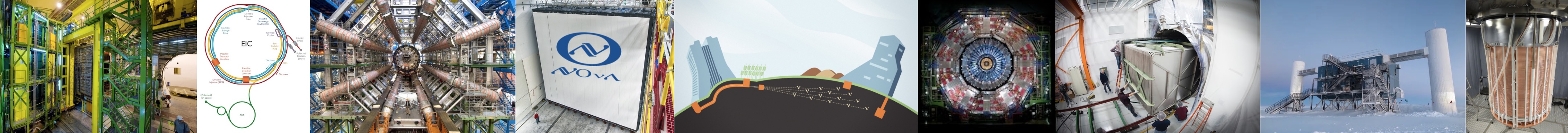}
    %\maketitle %this typesets the contents of \title, \author and \date
    \vskip 1.5cm
    \makeatletter
    {\LARGE \bfseries \@title \par}
    \vskip 1.0cm
    {\large \@date \par}
    \makeatother
    \vskip 1.5cm

    {\bf Abstract:}  Experimental particle physics seeks to understand the universe 
    by probing its fundamental particles and forces and exploring how they govern 
    the large-scale processes that shape cosmic evolution. This whitepaper presents 
    a vision for how Artificial Intelligence (AI) can accelerate discovery in this 
    field. We outline grand challenges that must be addressed to enable 
    transformative breakthroughs and describe how current and planned experimental 
    facilities can implement this vision to advance our understanding of the 
    vast and complex physical world from the smallest to the largest scales. We show how facilities currently under construction, such as 
    the HL-LHC, DUNE and soon EIC, can both benefit from and serve as proving grounds 
    for this vision, while also enabling a longer-term goal for how future experiments—
    like FCC-ee at CERN, IceCube-Gen2, a Muon Collider in the U.S., 
    and smaller to mid-scale projects—can be fully AI-native. 
    We describe how a truly national-scale collaboration, jointly managed across 
    large funding partners, and involving both DOE laboratories and universities, can 
    make this happen.
    
\vskip 0.7cm
    {\bf Audience:} The goal of this whitepaper is to highlight the emerging 
    opportunities and existing gaps for the broader particle physics community,
    to inform funding agencies in the event new resources become available, and to 
    highlight to policymakers the innovative contributions that experimental 
    particle physics can bring to the field. This document 
    is not intended as a 
    comprehensive proposal for all necessary activities, nor as an exhaustive 
    review of ongoing R\&D. 
    
\vskip 1.8in
\begin{tabulary}{1.0\textwidth}{L L}
\begin{minipage}[c]{0.1\textwidth}
    \centering
    \includegraphics[width=0.9\textwidth]{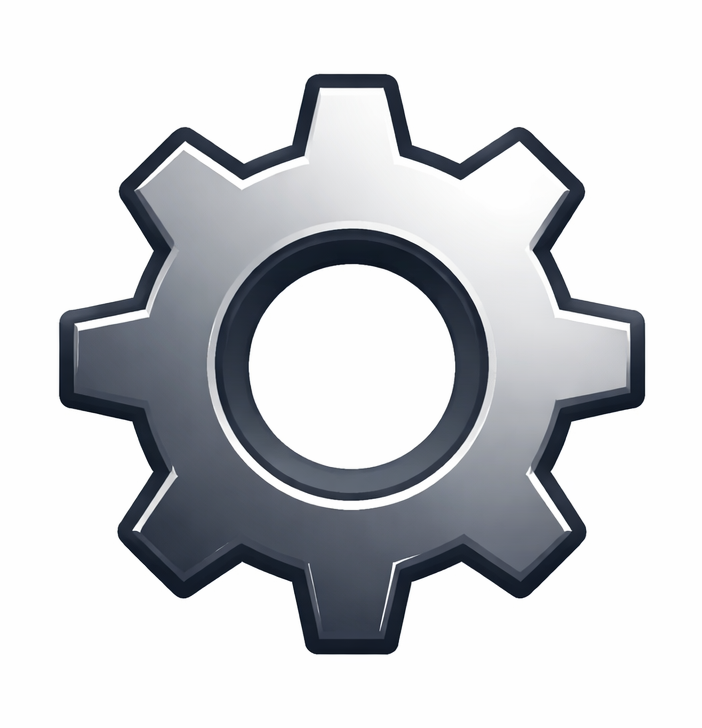} 
\end{minipage} 
&
\begin{minipage}[c]{\linewidth}
     This report has been produced in collaboration with, and is endorsed by, the Coordinating Panel for Software and Computing (CPSC) which is hosted by the Division of Particles and Fields (DPF) of the American Physical Society (APS).
\end{minipage} 
\end{tabulary}

}

\thispagestyle{empty}
\newpage

%\begin{center}
%{\bf SIGNERS/ENDORSERS}
%\end{center}
\begin{sloppypar}
\noindent Thea Klaeboe Aarrestad\textsuperscript{1}, 
Alaa Abdelhamid\textsuperscript{2}, 
Haider Abidi\textsuperscript{3}, 
Jahred Adelman\textsuperscript{5}, 
Jennifer Adelman-McCarthy\textsuperscript{6}, 
Shuchin Aeron\textsuperscript{7}, 
Garvita Agarwal\textsuperscript{8}, 
Usman Ali\textsuperscript{10}, 
Cristiano Alpigiani\textsuperscript{11}, 
Omar Alterkait\textsuperscript{7}
% ...
\end{sloppypar}
%\newpage
%\vspace{1em}
\vspace{0.5cm}

%\paragraph{Notes:} 
\noindent{\bf Notes:} 
{\small 
\roleEditor Editors, 
\roleWriter Primary text contributors
%\roleEndorser Endorser
}

%\vskip 0.5cm
%\begin{center}
%\small
%\input{institutions-print2.tex}
%\end{center}

%\begin{center}
%{\bf ENDORSERS}
%\end{center}
\vskip 0.5cm
\noindent The following additional people have endorsed the vision presented in this whitepaper:
Saleh Abubakar\textsuperscript{4}, 
Towsifa Akhter\textsuperscript{9}, 
Michael Albrow\textsuperscript{6}, 
Harut Avakian\textsuperscript{17}, 
Rachel Bartek\textsuperscript{22}, 
Steven Barwick\textsuperscript{24}, 
Jake Bennett\textsuperscript{32}, 
Antonio Boveia\textsuperscript{26}, 
Raymond Brock\textsuperscript{47}, 
David Caratelli\textsuperscript{51}, 
Doug Cowen\textsuperscript{62}, 
Jennet Dickinson\textsuperscript{71}, 
Christopher Dudley\textsuperscript{73}, 
James Eshelman\textsuperscript{75}, 
Yongbin Feng\textsuperscript{78}, 
Yuri Gershtein\textsuperscript{82}, 
Rebeca Gonzalez Suarez\textsuperscript{85}, 
Nate Grieser\textsuperscript{88}, 
Robert Hirosky\textsuperscript{74}, 
Doojin Kim\textsuperscript{108}, 
Doyeong Kim\textsuperscript{57}, 
Ryan Kim\textsuperscript{6}, 
Kyoungchul Kong\textsuperscript{111}, 
David Lawrence\textsuperscript{17}, 
Paul Lebrun\textsuperscript{6}, 
Chad Leino\textsuperscript{20}, 
Hong Ma\textsuperscript{3}, 
Samia Mahmood\textsuperscript{122}, 
Ritesh Pradhan\textsuperscript{141}, 
Vihanga Ranatunga\textsuperscript{18}, 
Ivan Razumov\textsuperscript{147}, 
Patrizia Rossi\textsuperscript{17}, 
Brooke Russell\textsuperscript{65}, 
Graziella Russo\textsuperscript{92}, 
Nihar Ranjan Saha\textsuperscript{106}, 
Heidi Schellman\textsuperscript{151}, 
Daniel Schulte\textsuperscript{90}, 
Ram Krishna Sharma\textsuperscript{106}, 
Wei Shi\textsuperscript{27}, 
James Simone\textsuperscript{6}, 
Aron Soha\textsuperscript{6}, 
Maciej Szymanski\textsuperscript{57}, 
Matt Toups\textsuperscript{6}, 
Emanuele Usai\textsuperscript{20}, 
Lian-Tao Wang\textsuperscript{34}, 
Jullian Watts\textsuperscript{2}, 
Jaehoon Yu\textsuperscript{53}

\vskip 0.5cm
\begin{center}
\small
\textsuperscript{1}ETH Zürich, 
\textsuperscript{2}University of Tennessee, Knoxville, 
\textsuperscript{3}Brookhaven National Laboratory, 
\textsuperscript{4}Erciyes University, 
\textsuperscript{5}Northern Illinois University, 
\textsuperscript{6}Fermi National Accelerator Laboratory, 
\textsuperscript{7}Tufts University, 
\textsuperscript{8}University of Notre Dame, 
\textsuperscript{9}Texas A\&M University, 
\textsuperscript{10}University of Oklahoma, 
\textsuperscript{11}University of Washington, 
\textsuperscript{12}Princeton University, 
\textsuperscript{13}IPM, 
\textsuperscript{14}Brandeis University, 
\textsuperscript{15}Lawrence Berkeley National Laboratory, 
\textsuperscript{16}Iowa State University, 
\textsuperscript{17}Thomas Jefferson National Accelerator Facility, 
\textsuperscript{18}Case Western Reserve University, 
\textsuperscript{19}Illinois Institute of Technology, 
\textsuperscript{20}University of Alabama, 
\textsuperscript{21}University of Minnesota, 
\textsuperscript{22}Catholic University of America, 
\textsuperscript{23}SLAC National Accelerator Laboratory, 
\textsuperscript{24}University of California, Irvine, 
\textsuperscript{25}Bandirma Onyedi Eylul University, 
\textsuperscript{26}Ohio State University, 
\textsuperscript{27}Stony Brook University, 
\textsuperscript{28}Siena University, 
\textsuperscript{29}University of Houston, 
\textsuperscript{30}Louisiana Tech, 
\textsuperscript{31}Brown University, 
\textsuperscript{32}University of Mississippi, 
\textsuperscript{33}Sejong University, 
\textsuperscript{34}University of Chicago, 
\textsuperscript{35}Southern Methodist University, 
\textsuperscript{36}Idaho State University, 
\textsuperscript{37}NERSC, Lawrence Berkeley National Laboratory, 
\textsuperscript{38}University of Iowa, 
\textsuperscript{39}University of Wisconsin–Madison, 
\textsuperscript{40}University of Nebraska-Lincoln, 
\textsuperscript{41}Morgridge Institute for Research, 
\textsuperscript{42}Northeastern University, 
\textsuperscript{43}University of Iowa and Marmara University, 
\textsuperscript{44}Texas A\&M University \& Hamad Bin Khalifa University, Qatar, 
\textsuperscript{45}University of Florida, 
\textsuperscript{46}Lancaster University, 
\textsuperscript{47}Michigan State University, 
\textsuperscript{48}Columbia University, 
\textsuperscript{49}The University of Kansas, 
\textsuperscript{50}Atlantis University, 
\textsuperscript{51}University of California, Santa Barbara, 
\textsuperscript{52}Universidad de Antioquia, 
\textsuperscript{53}University of Texas at Arlington, 
\textsuperscript{54}Harvard University, 
\textsuperscript{55}University of Warwick, 
\textsuperscript{56}University of Puerto Rico Mayaguez, 
\textsuperscript{57}Argonne National Laboratory, 
\textsuperscript{58}Vanderbilt University, 
\textsuperscript{59}University of New Hampshire, 
\textsuperscript{60}Illinois State University, 
\textsuperscript{61}University of Massachusetts Amherst, 
\textsuperscript{62}Penn State University, 
\textsuperscript{63}Carnegie Mellon University, 
\textsuperscript{64}ELTE Eotvos Lorand University, 
\textsuperscript{65}Massachusetts Institute of Technology, 
\textsuperscript{66}MIT and Harvard University, 
\textsuperscript{67}University of Tokyo, 
\textsuperscript{68}Yale University, 
\textsuperscript{69}Mississippi State University, 
\textsuperscript{70}University of Hawaii (Manoa), 
\textsuperscript{71}Cornell University, 
\textsuperscript{72}University of Manchester, 
\textsuperscript{73}University of Oregon, 
\textsuperscript{74}University of Virginia, 
\textsuperscript{75}Nova Software, Inc., 
\textsuperscript{76}Indiana University, 
\textsuperscript{77}William \& Mary, 
\textsuperscript{78}Texas Tech University, 
\textsuperscript{79}Simons Foundation, 
\textsuperscript{80}University of California, Berkeley, 
\textsuperscript{81}Stanford University, 
\textsuperscript{82}Rutgers University, 
\textsuperscript{83}Georgia Tech, 
\textsuperscript{84}University of South Alabama, 
\textsuperscript{85}Uppsala University, 
\textsuperscript{86}University of Minnesota Duluth, 
\textsuperscript{87}UC Berkeley/LBNL, 
\textsuperscript{88}University of Cincinnati, 
\textsuperscript{89}Johns Hopkins University, 
\textsuperscript{90}CERN, 
\textsuperscript{91}Oklahoma State University, 
\textsuperscript{92}University of California, Santa Cruz, 
\textsuperscript{93}Technical University of Munich, 
\textsuperscript{94}University of Colorado, Boulder, 
\textsuperscript{95}University of Washington / CENPA, 
\textsuperscript{96}University of Michigan, 
\textsuperscript{97}Rice University, 
\textsuperscript{98}University of Pittsburgh, 
\textsuperscript{99}University of Illinois Urbana-Champaign, 
\textsuperscript{100}UCSB / KITP, 
\textsuperscript{101}Virginia Tech, 
\textsuperscript{102}University at Buffalo, State University of New York, 
\textsuperscript{103}Rutherford Appleton Laboratory, 
\textsuperscript{104}Hampton University, 
\textsuperscript{105}New Mexico State University, 
\textsuperscript{106}Purdue University, 
\textsuperscript{107}Deutsches Elektronen-Synchrotron (DESY), 
\textsuperscript{108}University of South Dakota, 
\textsuperscript{109}Ohio University, 
\textsuperscript{110}Jeju National University, 
\textsuperscript{111}University of Kansas, 
\textsuperscript{112}Manhattan University, 
\textsuperscript{113}University of Texas at Austin, 
\textsuperscript{114}University of Wuppertal, 
\textsuperscript{115}Davidson College, 
\textsuperscript{116}University of California, Los Angeles, 
\textsuperscript{117}Northwestern University, 
\textsuperscript{118}iTHEMS/RIKEN, 
\textsuperscript{119}University of California, San Diego, 
\textsuperscript{120}Syracuse University, 
\textsuperscript{121}UCLouvain, 
\textsuperscript{122}Bellarmine University, 
\textsuperscript{123}Kansas State University, 
\textsuperscript{124}Christopher Newport University, 
\textsuperscript{125}Florida International University, 
\textsuperscript{126}University of Monastir - Tunisia, 
\textsuperscript{127}Polytechnic University of Madrid, 
\textsuperscript{128}Jawaharlal Nehru University, 
\textsuperscript{129}California Institute of Technology, 
\textsuperscript{130}A.I. Alikhanyan National Science Laboratory, 
\textsuperscript{131}University of Hyderabad, 
\textsuperscript{132}Colorado State University, 
\textsuperscript{133}Universidad Antonio Nariño, 
\textsuperscript{134}Niels Bohr Institute, 
\textsuperscript{135}The University of Edinburgh, 
\textsuperscript{136}IFIC-Valencia, 
\textsuperscript{137}Virginia Military Institute, 
\textsuperscript{138}University of Regina, 
\textsuperscript{139}University of Copenhagen, 
\textsuperscript{140}University of South Carolina, 
\textsuperscript{141}Indian Institute of Technology Hyderabad, 
\textsuperscript{142}University of Paris Saclay, 
\textsuperscript{143}FNAL/LPC, 
\textsuperscript{144}Duke University, 
\textsuperscript{145}University of Pennsylvania, 
\textsuperscript{146}Washington \& Jefferson College, 
\textsuperscript{147}Tomsk State University, 
\textsuperscript{148}Saint Louis University, 
\textsuperscript{149}IISER Tirupati, 
\textsuperscript{150}Florida State University, 
\textsuperscript{151}Oregon State University, 
\textsuperscript{152}Imperial College London, 
\textsuperscript{153}University Nebraska-Lincoln, 
\textsuperscript{154}IBS School, University of Science and Technology (UST), 
\textsuperscript{155}Washington College, 
\textsuperscript{156}Chung-Ang University, 
\textsuperscript{157}NSF IAIFI, 
\textsuperscript{158}Boston University, 
\textsuperscript{159}Islamic University of Madinah, 
\textsuperscript{160}The Niels Bohr Institute, University of Copenhagen, 
\textsuperscript{161}Irene-Joliot Curie Lab, 
\textsuperscript{162}University of Glasgow, 
\textsuperscript{163}Old Dominion University, 
\textsuperscript{164}University of Rochester, 
\textsuperscript{165}Princeton University/CERN, 
\textsuperscript{166}University of Freiburg, 
\textsuperscript{167}University of California, Riverside, 
\textsuperscript{168}University of Cambridge, 
\textsuperscript{169}TRIUMF, 
\textsuperscript{170}Lawrence Livermore National Laboratory, 
\textsuperscript{171}Nanjing Normal University, 
\textsuperscript{172}Fudan University, 
\textsuperscript{173}University of Parma, 
\textsuperscript{174}University of Maryland, College Park
\end{center}

\newpage

\tableofcontents
\newpage
\endgroup

\setcounter{page}{1}
\begin{center}
\includegraphics[width=1.0\textwidth, alt={LHCb, EIC, ATLAS, NOvA, DUNE, CMS, MicroBooNE, IceCube, XENON}]{images/aiml-banner.jpg}
\end{center}
    
\section*{Executive Summary}
\addcontentsline{toc}{section}{Executive Summary}
%(1 page)
\label{sec:executive-summary}
% Written last — punchiest hooks for non-HEP readers
% Key takeaways, vision statement, resource requirements at a glance, call to action
Experimental particle physics addresses some of the most fundamental questions 
about the universe through facilities that are among the largest, most complex, 
and ambitious scientific endeavors ever constructed. Across collider, neutrino, 
cosmic, and rare-event experiments, these facilities function as {\em massive and 
continuous data generators}, producing petabytes of rich, structured, curated 
data annually, while discarding a majority of the raw information due to 
bandwidth, storage, or latency constraints.  The scale, complexity, and structure 
of these datasets align with the strengths of modern Artificial Intelligence 
(AI): high-dimensional pattern recognition, rare-signal inference, low-latency 
decision making, and the orchestration of complex systems spanning hardware, 
software, and human expertise.  
AI can play a transformative role by enabling experiments to extract and retain more information from data, extending the discovery potential, 
and reducing the time from data-taking to discovery. It can also improve the efficiency and sustainability of long-running facilities and increase sensitivity to subtle or unexpected phenomena.
Now is a pivotal moment: experiments currently 
in operation or under construction will define the scientific output of particle 
physics for the next several decades, while unprecedented national investments in 
AI, advanced computing, and workforce development create a rare opportunity to 
couple our scientific challenges with foundational AI research.  This whitepaper 
presents a community vision and an actionable plan to seize this moment.

\begin{wrapfigure}{r}{0.47\textwidth}
\centering
\vskip -7.0mm
\includegraphics[width=1.00\linewidth]{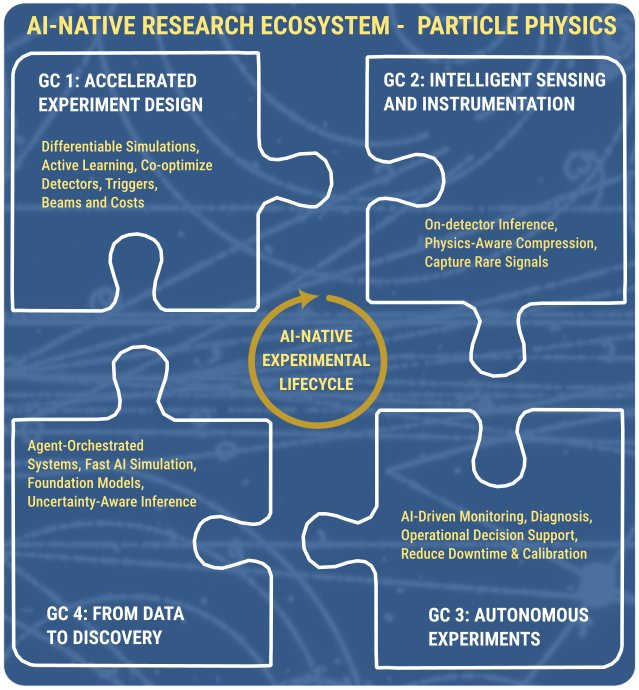}
\caption{Grand Challenges spanning the experimental lifecycle in particle physics.}
\vskip -4.0mm
\label{fig:codashep}
\end{wrapfigure}

Our vision is to embed AI end-to-end across the experimental lifecycle,
from the co-design of accelerators and detectors to intelligent sensing, data 
acquisition, autonomous operations and calibration, and accelerated analysis for 
discovery.  In this “AI-Native” paradigm, experiments become continuously 
learning systems: design and operations are optimized over time, 
more information 
is retained and understood, and scientists spend less time on mechanical steps 
and more on scientific interpretation - directly advancing the national 
goal of dramatically increasing research output and discovery. This 
approach also maximizes the return on current investments and enables 
next-generation facilities to be conceived and operated with AI as 
a core capability. 

\vskip 0.1cm
\noindent{\bf Grand Challenges:} This vision is organized around four Grand Challenges that form a
self-reinforcing engine for an AI-Native ecosystem. 
%{\it Accelerated Experimental Design} 
{\color{darkgray}\textbf{\textit{Accelerated Experimental Design}}}
uses differentiable, agent-guided optimization to co-design 
accelerators, and detectors so that ambitious ideas become buildable, 
higher-impact experiments 
on faster timescales with reduced technical risk and costs. 
%{\it Intelligent Sensing \& Instrumentation}
{\color{darkgray}\textbf{\textit{Intelligent Sensing \& Instrumentation}}}
moves intelligence upstream via trigger-less or 
AI-assisted readout, physics-aware compression, and real-time inference to 
preserve rare, unexpected, or time-critical signals while operating within 
bandwidth and storage constraints. 
%{\it Autonomous Experiments} 
{\color{darkgray}\textbf{\textit{Autonomous Experiments}}}
transforms 
labor-intensive, reactive operations into proactive, resilient, and continuously 
calibrated systems, capturing more high-quality data with less downtime and 
preserving institutional knowledge over decades-long experiment lifetimes. 
%{\it From Data to Discovery}
{\color{darkgray}\textbf{\textit{From Data to Discovery}}}
integrates foundation models, fast 
AI-enabled reconstruction and simulation, and agent-orchestrated workflows to 
compress analysis cycles by orders of magnitude and open new regions of theory 
space to exploration. Advancing any one challenge amplifies the others, expanding 
scientific reach and productivity while positioning the U.S. as a global 
leader in AI-powered particle physics.

\vskip 0.1cm
\noindent{\bf National-Scale Collaboration:} To realize this vision, we propose a {\em national-scale collaboration} 
that brings together DOE national laboratories, U.S. universities, and 
industry partners to 
develop and deploy shared AI capabilities, impact multiple 
experiments, and train an AI-literate scientific workforce.
Experimental particle physics is unique among scientific fields in its 
ability to leverage decades of experience building and managing national-scale 
collaborations.
Modeled on successful U.S. operations programs, we envision a national
core effort of 120 FTEs or larger, to be jointly managed over multiple 
large funding partners, and guided by Grand Challenges to 
build an AI-native research ecosystem. 
%this effort will coordinate execution across the community. 
An explicit collaboration avoids duplication while accelerating discovery and technological impact.
DOE national laboratories would provide scale, advanced 
computing resources, and long-term stewardship of complex projects, while 
universities would drive innovative R\&D and workforce development through deep 
engagement of students and postdocs. NSF's mission and community 
would allow expansion beyond particle physics and
provide pathways to engage other domain sciences and translate the impact
more broadly in  the U.S. research and education ecosystem. 
Durable impact requires university partnerships structured at a scale comparable to lab programs.
This flexible structure also enables targeted investments by private 
philanthropic partners in innovation, education, and technology translation.
Strategic partnerships with 
industry would ensure rapid access to state-of the-art AI technologies and best 
practices, enabling efficient technology transfer and co-design. Together, this 
collaboration would create a cohesive national capability that strengthens 
connections between labs, universities, and industry and translates U.S. 
investments in AI and particle physics into sustained scientific leadership and 
workforce impact.

\vskip 0.1cm
\noindent{\bf Workforce Development:} The national-scale collaboration we
propose would have connections to nearly all U.S. Ph.D.-granting 
universities and many undergraduate institutions. The DOE Genesis Mission 
has a goal of 
training 100,000 scientists over the next decade. 
Workforce development activities integrated with such a
collaboration could contribute 2100 PhDs and roughly 20,000 undergraduates
to that goal, with potential for more via neighboring fields such as nuclear physics. 

\vskip 0.1cm
\noindent{\bf Leveraging National Infrastructure:} Our vision is built on an
integrated ecosystem in which advanced AI technologies and national-scale
cyberinfrastructure jointly support the full experimental lifecycle.
Physics-aware AI capabilities such as multimodal foundation models, fast and
differentiable simulation with simulation-based inference, and agentic active
learning systems coordinate complex scientific workflows.
Delivered through inference-as-a-service and embodied in high-fidelity digital
twins, these tools link design, operations, and analysis into a closed loop,
transforming particle physics facilities into continuously learning systems and
dramatically accelerating the path from data to discovery. Realizing this vision
requires a new generation of AI-native cyberinfrastructure that extends the
community’s strengths in distributed computing and data management. It
must support large-scale model training, always-on low-latency
inference for real-time operations, and AI-accelerated simulation tightly
integrated with High Performance Computing facilities. Central to this effort is
a federated, AI-ready data ecosystem that follows Findable, Accessible,
Interoperable, and Reusable (FAIR) principles and enables cross-experiment
learning while respecting governance and ownership. Leveraging national assets
such as the DOE Leadership Class Facilities, the emerging American Science
Cloud, and partnerships with NSF, universities, and industry, this ecosystem
provides shared training, inference, workflow orchestration, and interoperable
data platforms, amplifying the scientific return of U.S. investments in particle
physics and AI.

\newpage

\section{Science Drivers}
%(1 page)
\label{sec:science-drivers}
% [1 page] - Points to P5/ESPPU for details
% [Sarah, Tulika]

Particle physics experiments include some of the largest and most complex scientific endeavors. Thousands of physicists, engineers, technicians and students work together to build and operate facilities producing some of the largest datasets in the world in order to unlock answers to foundational questions about the universe. Our experiments already record data samples of 100s of petabytes of data annually, resulting in datasets that are multiple exabytes in size to be analyzed. In addition, the fraction of data that is currently stored and processed can be much less than one percent of the total data generated. Overcoming these limitations due to data transfer and available computing capacity presents a great opportunity for scientific expansion. Experimental particle physics is uniquely positioned to unleash the potential of the AI revolution for science at scales where these methods are transformational. 

The Particle Physics community in the U.S. sets its scientific priorities through the Snowmass community process that serves as input to the Particle Physics Project Prioritization Panel (P5). The most recent P5 report~\cite{hep_p5_2023}, from 2023, articulates a compelling scientific vision for U.S. particle physics over the coming decade, focused on a number of foundational questions about the universe. These science drivers span the smallest known scales and the largest cosmic structures, motivating an experimental program of unprecedented scale, complexity, and duration. P5 prioritizes maximizing discovery from major initiatives such as the High Luminosity LHC (HL-LHC), Deep Underground Neutrino Experiment (DUNE) and Proton-Improvement Plan II (PIP II), and Legacy Survey of Space and Time (LSST) at the Rubin Observatory, while advancing critical R\&D toward a future Higgs factory, and ultimately a 10~TeV parton center-of-momentum (pCM) collider capable of probing new physics at unprecedented energies.  

{\bf Deciphering the Quantum Realm: Higgs Physics and New Phenomena}
A central P5 priority is to use the Higgs boson as a precision tool to probe fundamental physics scales. Direct searches for new particles and interactions and indirect tests via quantum imprints of new phenomena leverage the HL-LHC program and motivate future collider facilities. These measurements demand exquisite control of detector performance, enormous simulated datasets, and sophisticated statistical inference across vast theory spaces. AI can accelerate this program by enabling automated, optimizable analyses; rapid-turnaround simulation and reconstruction; and global searches for subtle deviations from Standard Model predictions. By reducing analysis latency and expanding the accessible parameter space, AI techniques can directly increase the discovery reach of flagship collider experiments and play a central role in the design of future experiments and facilities.

{\bf Deciphering the Quantum Realm: Elucidating the Mysteries of Neutrinos}
Understanding the nature of neutrinos -- their mass ordering and values, as well as role in matter–antimatter asymmetry, and (astro)physical processes -- is a top P5 science driver. Long-baseline experiments such as DUNE and a rich portfolio of complementary measurements from experiments such as NOvA, T2K, the Short-Baseline Neutrino Program (SBN), and IceCube are positioned to address these fundamental questions. Neutrino experiments are characterized by low effective signal rates for key physics channels, complex detector responses, and long operational lifetimes. AI-enabled reconstruction, calibration, and data quality monitoring can substantially improve signal efficiency and background rejection, while autonomous operations can increase data taking efficiency and data quality over decades. These gains would translate directly into faster and more precise answers to some of the most fundamental open questions in particle physics.

{\bf Pursue quantum imprints of new phenomena (flavor and precision frontier)}
The P5 report highlights precision measurements of rare processes as a powerful path to uncover new physics beyond the Standard Model. Belle-II and LHCb form the backbone of this program, delivering complementary sensitivity to rare decays, CP violation, and tests of lepton flavor universality in the beauty, charm, and tau-lepton sectors. AI is increasingly essential to this effort, enabling more efficient triggering, improved event reconstruction, background suppression, and global analyses that jointly control statistical and systematic uncertainties across many channels. In the muon sector, the Mu2e experiment will search for coherent muon-to-electron conversion in the field of a nucleus with unprecedented sensitivity, providing a decisive probe of charged‑lepton‑flavor violation. AI techniques can enhance signal discrimination, suppress backgrounds from cosmic rays and beam‑related processes, and optimize operations and calibration over long running periods. Looking ahead, the P5 report emphasizes R\&D toward an advanced muon facility. AI will be a critical tool for ultra-low background operation and rare event searches across the muon program. Together, these experiments exemplify how precision measurements, accelerated by AI, has the potential to reveal new fundamental physics through subtle quantum effects.

{\bf Illuminating the Hidden Universe: Dark Matter, Cosmic Evolution, and Beyond ``Traditional'' Astronomy}
The P5 report emphasizes a diverse experimental strategy to determine the nature of dark matter, to understand what drives cosmic evolution, including dark energy, inflation, and structure formation, and some of the most cataclysmic processes in the universe. This portfolio spans terrestrial direct‑detection and axion experiments (e.g., DarkSide‑20k, LZ, SuperCDMS, XENONnT, ADMX), accelerator‑based probes, and astrophysical/cosmological observations. IceCube anchors the multi‑messenger neutrino program and, together with next‑generation upgrades (e.g., IceCube‑Gen2), enables indirect dark‑matter searches and real‑time studies of cosmic accelerators and transients.  AI is uniquely suited to this overall landscape: it enables anomaly detection beyond predefined signal models, rapid cross correlation of disparate data streams, and real-time responses to transient cosmic events. In this way, AI can expand sensitivity not only to well motivated theories, but also to the unexpected. Dark energy observatories share many common challenges in data volume, complexity, and analysis, as described in a dedicated LSST and Dark Energy Science Collaboration (DESC) AI/ML white paper~\cite{lsst-desc}. This document concentrates on experimental particle physics facilities and science drivers, while recognizing strong opportunities for cross‑fertilization of AI tools, infrastructure, and best practices across these domains. 

% Question how IceCube fits in here. There is the argument about 1) neutrinos as probes of astrophysical phenomena, 2) neutrinos as probes the highest energy, and furthest away portions of the universe, 3) neutrinos as drivers of fundamental astrophysical pehnomena (CCSN). We can split a "neutrino measurements" with IceCube Upgrade and a "multi-messenger" Tulika: added a couple of sentences for IceCube abd IceCube-Gen2. 

{\bf Nuclear Science Drivers: Understanding Visible Matter and Probing Lepton Number Violation} The 2023 Long Range Plan for Nuclear Science (LRP)~\cite{osti_2280968} complements the P5 vision by elevating the Electron Ion Collider (EIC) and a multi-isotope neutrinoless double beta decay campaign as priorities to understand how visible matter is built and whether neutrinos are their own antiparticles. 
The primary science driver for the EIC is to understand how the strong force
gives rise to the mass, spin, and internal structure of visible matter. It will
map the distributions and dynamics of quarks and gluons inside protons,
neutrons, and nuclei with unprecedented precision. 
The EIC has the potential to become the first major facility to incorporate AI
at all stages of design, construction, and operation. The primary experiment at
the EIC, ePIC, features a compute-detector integration that enables seamless
data processing from detector readout through physics analysis, with the goal of
enhancing scientific precision and accelerating the path to discovery. This
integration is based on streaming readout and AI driven workflows, and it
requires the implementation of AI algorithms to support adaptive, low latency,
data driven decision making at a scale and complexity beyond those of
traditional approaches.
The EIC is naturally aligned
with AI-driven reconstruction and analysis because its physics program relies on
fully differential, high rate measurements across complex final states. 
%AI has the potential to enable real-time data reduction, precision multidimensional
%unfolding, and global correlation analyses that are intractable with traditional
%methods. In addition, the integration of the Data Acquisition (DAQ) and offline
%computing requires the implementation of AI algorithms to enable adaptive,
%low-latency, data-driven decision making at a scale and complexity beyond those
%of traditional methods.

Both the P5 report and the LRP recognize neutrinoless double-beta decay as an important ``quantum imprint'' measurement that complements neutrino oscillation experiments and collider searches. It seeks to reveal whether neutrinos are their own antiparticles and whether lepton number is violated; answers with far reaching implications for the origin of mass and the matter–antimatter asymmetry of the universe. Because the signal is extraordinarily rare, backgrounds must be controlled at unprecedented levels. Accelerating this drive for precision discovery in ultra rare measurements would particularly benefit from the potential use of AI in analysis and operations.
%, directly aligning with this whitepaper's vision for accelerating discovery in ultra rare, precision measurements.

In the years preceding the start of the EIC physics program, the 2023 LRP
emphasizes the completion of the RHIC mission and its transition into the EIC,
while noting that the LHC heavy-ion program will provide the world-leading
facility for hot-QCD studies in that pre-EIC era. At the LHC, the ALICE
experiment is dedicated to studying the quark-gluon plasma with properties of
the early universe, at temperatures of trillions of degrees and with near
symmetry between matter and antimatter. A deep AI integration in the
next-generation ALICE 3 detector, designed to replace ALICE and operate in 2035
and beyond, would be transformational in achieving and extending the
experiment’s physics output. Similarly, medium-energy efforts at Jefferson
Lab—including the Measurement of a Lepton-Lepton Electroweak Reaction
(MOLLER)~\cite{moller-jlab} and the Solenoidal Large Intensity Device
(SoLID)~\cite{solid-jlab} experiments within the ongoing CEBAF 12 GeV
program—will benefit from further integration of emerging AI technology across
detector design, real-time operation, reconstruction, and uncertainty-aware
analysis. Looking ahead, the LRP also highlights a staged CEBAF evolution that
begins with high-duty-cycle polarized positron capabilities at 12 GeV and could
progress to a \(\sim\)22 GeV electron program while maintaining world-leading
luminosity; fully exploiting these opportunities will rely on AI-enabled tooling
spanning all four Grand Challenges.

{\bf Enabling the Long-Term Vision: Future Facilities and Sustainability}
Both the P5 report and a recent National Academy of Sciences~\cite{NASEM2025HiggsBeyond28839} report underscore the importance of pursuing an ambitious long-term vision.
%, including next generation collider concepts and next-generation multi-messenger astrophysics facilities, while maximizing the scientific return of current investments.  
A Higgs factory such as the FCC-ee will be a crucial step toward fully revealing
the secrets of the Higgs boson within the quantum realm and will be a sensitive
probe of the quantum imprints of new phenomena. Proposed multi-messenger
facilities, like IceCube-Gen2, will probe the furthest and most energetic
corners of the universe, while also showing how neutrinos contribute to cosmic
mechanisms. On a longer timescale, a 10~TeV pCM collider such as a muon collider
potentially sited in the U.S., FCC-hh, or a wakefield-based $e^{+}e^{–}$ collider would
enable a comprehensive physics portfolio that includes ultimate measurements in
the Higgs sector and a broad search program. In each case, AI-driven design and
optimization can reduce technical risk, improve cost and schedule realism, and
enable holistic co-design of accelerators, detectors, and computing
infrastructure. AI can make large experiments more sustainable by reducing labor
intensive operations and preserving institutional knowledge, both critical
factors for projects spanning multiple decades.

Finally, small-scale experiments play a critical role in our scientific
ecosystem and are a natural focus area for AI-enabled innovation. These
experiments, often with targeted physics goals and short lifecycles, offer
exceptional agility for deploying and validating new AI approaches in
reconstruction, simulation, calibration, operations, and analysis. Within the
Advancing Science and Technology through Agile Experiments (ASTAE) program
highlighted by P5, AI can dramatically accelerate experimental design, reduce
labor-intensive operations, and enhance sensitivity in resource-constrained
environments. Moreover, successful AI methodologies developed and demonstrated
in these projects can be transferred to major facilities, amplifying their
impact while minimizing risk. 

\newpage

\section{Vision: AI-Native Particle Physics}
\label{sec:vision}
%[Tulika, Peter, Verena]

Our science drivers define what we seek to understand about the universe. Facilities and techniques define how rapidly, broadly, and effectively we can pursue those goals. As experimental particle physics enters an era of unprecedented scale and complexity, AI has become a foundational capability essential to realizing the full scientific potential of our facilities.  We envision a future in which AI is embedded end-to-end across the entire experimental lifecycle -- from the design and optimization of future facilities, through intelligent sensing and autonomous operations, to simulation, data processing, analysis, and scientific discovery.  In this vision, AI is not merely a supporting tool, but a strategic accelerator of scientific impact: managing complexity at unprecedented scales, enabling leaps in productivity, dramatically shortening time to insight, significantly increasing experimental uptime and efficiency, and opening pathways to new transformative scientific capabilities. In short, AI becomes a cornerstone of the future of experimental particle physics and a prerequisite for maintaining U.S. leadership in data-intensive science. 

\begin{figure}[!htbp]
    \centering
    \includegraphics[width=0.7\linewidth]{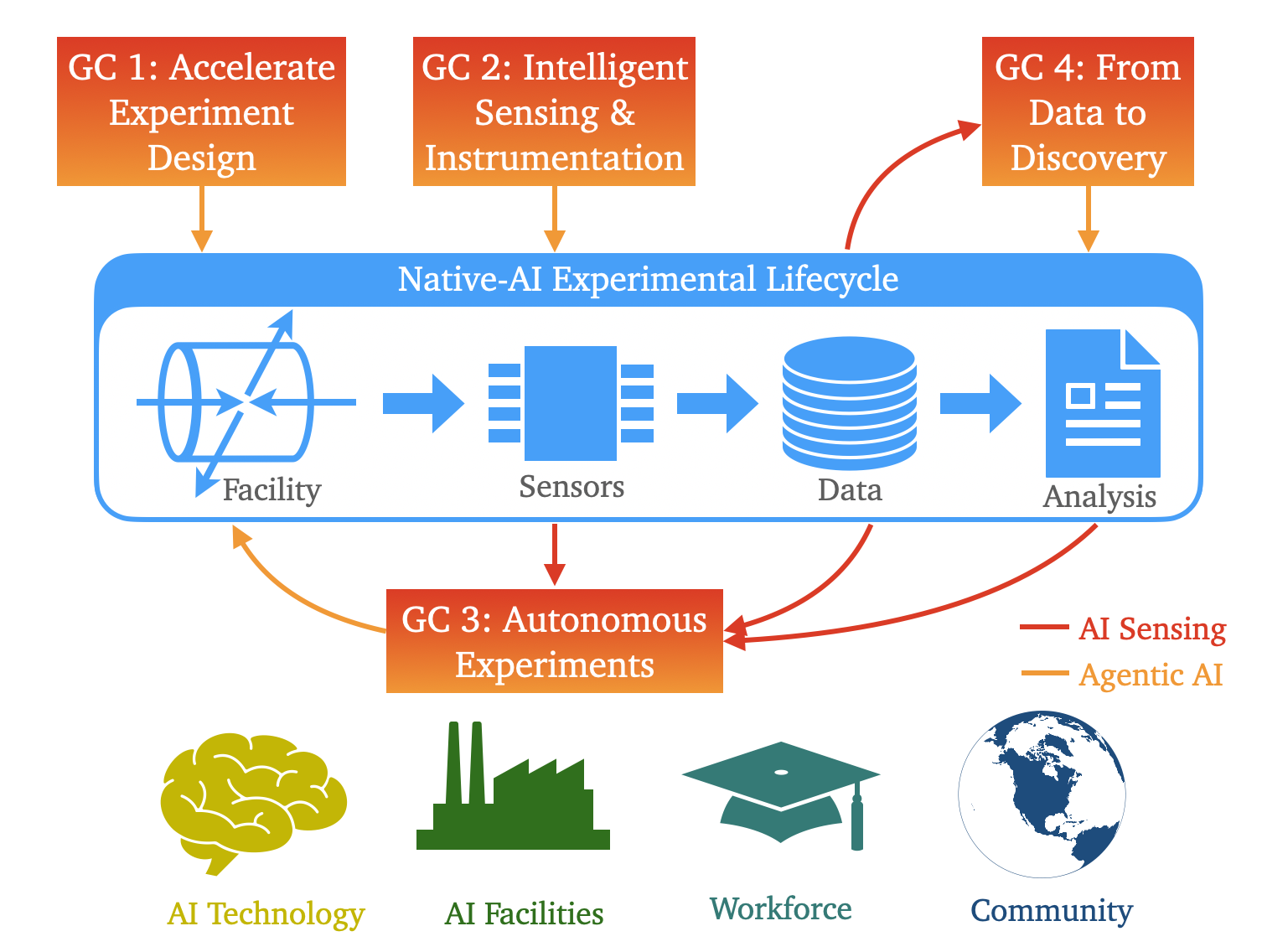}
    \caption{Diagram of the AI-Native experimental lifecycle including the facility/detector, sensing/instrumentation, data acquisition/curation, and analysis. The highlighted Grand Challenges identify where transformative advances enabled by AI accelerate the realization of this AI‑Native vision. The lower panel illustrates the shared technologies and infrastructure that underpin and sustain the ecosystem. }
    \label{fig:GCSummary}
\end{figure}

To make this vision actionable, we articulate four Grand Challenges that together cut across current and future particle physics experiments. As summarized in Fig.~\ref{fig:GCSummary}, these challenges are: 
\begin{itemize}
    \item \textbf{Grand Challenge 1: Accelerated Experimental Design} - Use of differentiable/surrogate simulations, and active learning to co-optimize detectors, triggers, beams, and costs, informing design choices for future experimental facilities.
    \item \textbf{Grand Challenge 2: Intelligent Sensing and Instrumentation} - Moving intelligence upstream with on-detector inference, trigger-less/AI-assisted readout, and physics-aware compression to capture rare or unexpected signals without overwhelming bandwidth or storage.
    \item \textbf{Grand Challenge 3: Autonomous Experiments} - Automating facility operations and calibration with AI-driven monitoring, diagnosis, and operational decision support to reduce downtime, shorten calibration cycles, and preserve institutional knowledge.
    \item \textbf{Grand Challenge 4: From Data to Discovery} - Building agent-orchestrated, goal-directed analysis systems that integrate foundation models, AI-accelerated reconstruction and simulation, and uncertainty-aware inference to dramatically reduce analysis latency, increase scientific productivity, and expand discovery reach.
\end{itemize}

Advancing any one Grand Challenge boosts the others (e.g., faster simulation results in better design and search reach), creating a self-reinforcing pipeline that improves sensitivity, reduces latency, and increases returns across the science portfolio. Together these Grand Challenges provide a natural organizing principle for a 
national-scale, multi-institutional collaboration that unites universities, laboratories, and industry partners. Shared capabilities, reusable infrastructure and cross-cutting expertise will enable such a collaborative effort to integrate these novel pipelines into scientific best practices used by experiments.
The community can then rally around a unifying goal: enhancing scientific methodology with AI in facilities across experimental particle physics.
%to deliver immediate benefits at flagship facilities, while using small- and mid-scale agile experiments as rapid testbeds that transfer successful approaches to the flagships. 

In our vision, the design and operation of experiments is fundamentally transformed by advances in AI. In an ``AI-Native'' paradigm, facilities, data analysis, simulations, and operations are co-designed with intelligence from the outset, enabling continuously optimized performance, resilient and adaptive operations, and inference that tightly connects data to theory. 
Near-term demonstrators establish these capabilities at scale; mid-term efforts translate them to standard practice; enabling the next generation of particle physics experiments to be fully AI-native facilities. Demonstrators at flagship facilities can target parts of the experimental lifecycle: HL-LHC and LBNF/DUNE operations; FCC-ee design studies, IceCube-Gen2 deployment, or Muon Collider R\&D; while small- and mid-scale agile experiments could be a basis for rapid testbeds covering the entire experimental lifecycle.
%Near-term demonstrators at the HL-LHC and DUNE establish these capabilities at scale; mid-term efforts such as FCC-ee design studies, IceCube-Gen2 deployment, and Muon Collider R\&D translate them into standard practice; and the long-term outcome is a new generation of fully AI-native flagship facilities. 
This staged and deliberate trajectory reflects the community's ambition and aligns with P5 guidance to maximize the scientific return of today’s investments while building the technological foundation for transformative, next-generation experiments that will define the future of particle physics.

\newpage
\section{Grand Challenges}
\label{sec:grand-challenges}

%We have organized the many community
%inputs we collected for this white-paper
%into the following four Grand Challenges. 

In this document we are framing our vision around a set of {\em Grand Challenges},
rather than individual projects.
%provides a unifying direction for the field. 
The Grand Challenge approach is better matched 
to developing long-term road-maps with shared benchmarks and 
infrastructure, focuses R\&D on real-world impact, and encourages 
both collaboration and healthy competition. It also raises the visibility
of the effort to attract talent from within and beyond the 
field. Most importantly, it creates a compelling narrative to engage 
the community and  potential resource providers.

% In this section, we have attempted to organize the many community inputs
% into a concise set of Grand Challenges. Framing our vision 
% around these challenges, rather than individual projects, provides a 
% unifying direction for the field. This approach is better matched 
% to developing long-term roadmaps with shared benchmarks and 
% infrastructure, focuses R\&D on real-world impact, and encourages 
% both collaboration and healthy competition. It also raises the visibility
% of the effort to attract talent from within and beyond the 
% field. Most importantly, it creates a compelling narrative to engage 
% the community and  potential resource providers.

In preparing the following list of Grand Challenges, we build on several recent community efforts, 
including the report of the Computational Frontier Topical Group on Machine 
Learning from the Snowmass 2021 community planning 
process~\cite{shanahan2022snowmass2021computationalfrontier}. 
More recently, the
Accelerated AI Algorithms for Data-Driven Discovery (A3D3) institute
hosted a workshop on AI to Accelerate Science and Engineering Discovery in 
October 2023 and produced a corresponding report~\cite{ai2ased}.
Similar topics were developed by the NSF Institute for Artificial Intelligence 
and Fundamental Interactions (IAIFI),
which produced a report~\cite{iaifi2025} on Generative AI and Discovery in 
the Physical Sciences. There was also a community 
report from the NSF Workshop on the Future of Artificial Intelligence and the 
Mathematical and Physical Sciences (MPS) held in March 2025~\cite{nsf-mps-2025}.
We also note the 
emergence of curated community resources on the use of machine learning in 
high-energy physics, such as the High Energy Physics (HEP) ML Living 
Review~\cite{hepml-living-review-website, feickert2021livingreviewmachinelearning},  
as well as focused resources on 
specific topics including simulation-based inference~\cite{sbi-website}.
As this is a fast-moving area, we also requested inputs from the 
particle physics community regarding this specific vision document 
through multiple forums organized by the APS Division of Particles and Fields 
(APS DPF). The aim of this request for input was to assemble a ``big 
picture'' view of the community vision rather than an exhaustive review
of all ongoing AI/ML activities and projects.
Between December 2025 and January 2026, we received more than 100 
contributions from roughly 150 individuals representing 50 
institutions (7 national laboratories and 43 universities), 
outlining opportunities for AI/ML to advance particle physics. The
community has significant interest in this area: it is clear that
a longer period for input would have resulted in many more contributions
from an even greater number of colleagues from across the field.
That said, the submissions spanned a broad range, from highly targeted 
project ideas to more ambitious strategic visions and represent
a sampling of the field.
Some focused on opportunities or gaps in
specific experiments, subfields (such as collider or neutrino physics), 
physics topics, or detector technologies, while others emphasized 
opportunities with impact across the field as a whole. 
The breadth of these contributions and the many efforts in recent years, 
together with the depth of expertise they reflect, underscore the 
community’s strong potential to leverage AI/ML for transformational 
advances in experimental particle physics.

%\begin{enumerate}
%    \item \textbf{Create a unifying vision}
%    \item \textbf{Motivate interdisciplinary collaboration}
%    \item \textbf{Attract funding and political support}
%    \item \textbf{Increase visibility and attracts talent}
%    \item \textbf{Provide a compelling narrative for success}
%    \item \textbf{Encourage competition and cooperation}
%\end{enumerate}

\subsection{Grand Challenge 1: Accelerated Experimental Design} 
%(1 page)}
%(e.g. neutrino detectors,  muon collider design and feasibility demonstrated rapidly?)[Sergo, Tova, Terao, Kagan]
%[Reviewers: Heather ]

The main goal of this challenge is to inform design choices for future experimental facilities through the use of differentiable/surrogate simulations and active learning to co-optimize detectors, triggers, beams, and costs.

Particle and nuclear experiment facility design requires years of dedicated expert intuition to tune thousands to millions of often highly interdependent parameters, which can typically only be achieved through slow iterations and greedy component-wise optimization. AI-based design and optimization presents a new avenue to address these critical challenges of scale and complexity and, crucially, navigating this complexity may lead physicists to new and unexpected designs. Future experimental facilities in particular, which are unconstrained by existing hardware and operations, present unique opportunities for AI-powered conceptual design and technology development. 

\begin{figure}[!htbp]
    \centering
    \includegraphics[width=0.90\linewidth]{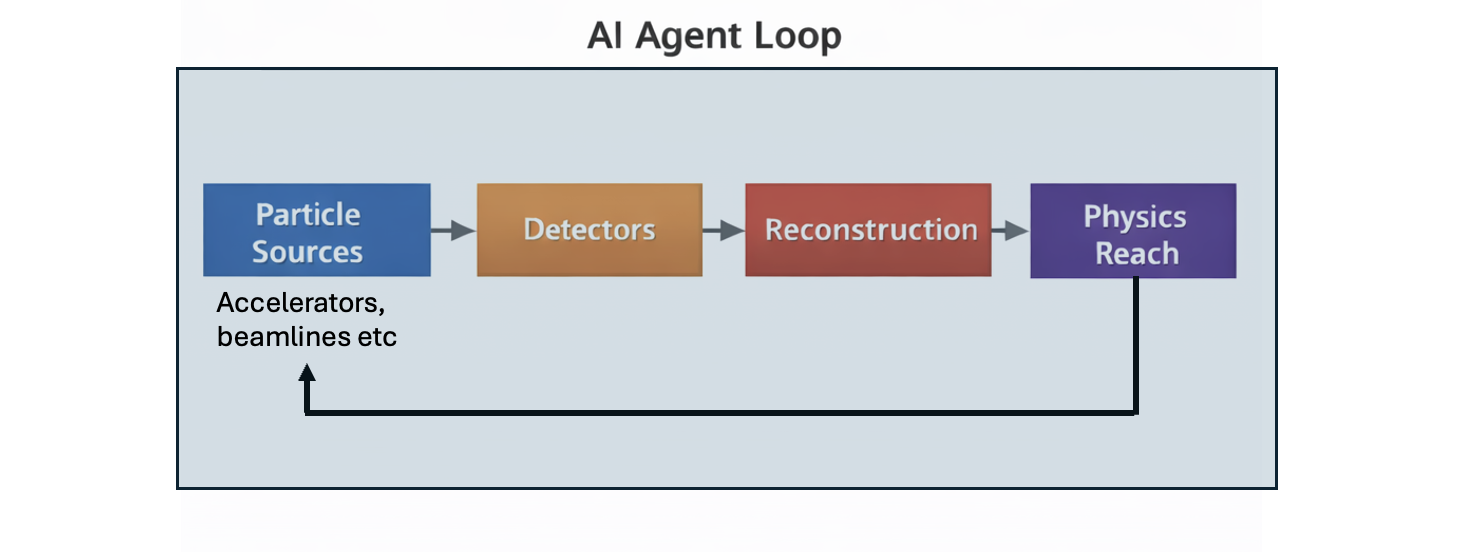}
    \caption{Conceptual schematic of an AI-driven design and optimization loop for future particle physics facilities. }
    \label{fig:AIAgentLoop}
\end{figure}

AI-driven optimization enables the simultaneous exploration of many accelerator and detector design spaces, capturing complex interdependencies that are difficult to address with traditional methods. Detector layout and design can be optimized together with machine parameters, enabling detector performance metrics or even physics analysis capabilities to enter directly into end-to-end optimization loops. This AI-native approach is illustrated in Fig.~\ref{fig:AIAgentLoop} Technologies such as high-power targetry, high-field magnets based on high-temperature superconductors, and advanced radio frequency acceleration systems can be similarly optimized within a unified and differentiable framework.

For the most technically mature large-scale future projects, AI-based systems
can re-optimize beam configurations and detector layouts to improve experimental sensitivity and technical robustness. As an example, at LBNF/DUNE Phase II, beamline, target and detector configurations involve a high-dimensional space with huge impact on the physics reach of the experiment. At FCC-ee, prototype detector geometry and technology, as well as reconstruction techniques, must still be optimized within the framework of an evolving accelerator design to reach the ambitious precision level for achieving the physics goals. AI-driven approaches to these optimizations provide medium-term, high-impact applications of AI that build directly on existing simulation and design workflows. 

In parallel, significant opportunities exist for applying AI-driven methods to accelerator physics, complementing the experimental design focus described above. The Accelerator and Beam Physics thrust of the
General Accelerator Research and Development (GARD) program in DOE has produced a roadmap~\cite{beamroadmap} that identifies a set of key areas, including beam intensity, beam quality, beam control, and predictive virtual accelerators, that increasingly rely on advanced AI techniques. Across these areas, AI-enabled surrogate modeling, real-time optimization, and predictive digital twins can transform how accelerators are designed, commissioned, and operated, enabling higher intensities, improved beam quality, and more precise control than is achievable with traditional approaches. Integrating these accelerator-focused AI advances with experiment-facing optimization frameworks strengthens the end-to-end vision for future facilities and ensures coherence between accelerator performance, detector design, and physics reach.

The R\&D path towards a future muon collider, recommended by both the P5 and
National Academies EPP reports, could provide the ultimate application of this
approach, with the potential to develop the first AI-native experimental
facility. In contrast to DUNE and FCC-ee, where several important design
elements remain under development but the overall concepts are well established, many key technologies for a muon collider have yet to be fully designed or demonstrated, and the ultimate physics reach and reliability of the facility depends on strong and non-trivial correlation among its many subsystems. Accelerator design, beam dynamics, machine-detector interface, and detector layout are each dictated by millions of interconnected parameters which should be holistically optimized to achieve viable and competitive performance. This approach could similarly benefit other proposed 10 TeV pCM machines, such as FCC-hh or a potential wakefield machine. 

Smaller and mid-scale agile experiments with targeted physics programs,
including those highlighted in the P5 report under the ASTAE program, provide
an abundance of
similar opportunities, and often share personnel with the large-scale
facilities. These experiments can directly benefit from AI tools, workflows, and
best practices developed within larger collaborations, enabling performance
optimization, faster design iteration, and improved cost and schedule efficiency
at a scale appropriate to their scope. They can also serve as smaller-scale
test-beds for new approaches, which can then be adopted by the large
collaborations. The transfer of AI enabled methodologies across experimental
programs amplifies their overall impact and helps build a more coherent and
sustainable ecosystem for future facilities. An example of such effort is
PIONEER~\cite{pioneercollaboration2025europeanstrategyparticlephysics}, a small
scale high-rate experiment to study the rarest decays of the charged pion. At
the heart of the proposed apparatus, PIONEER features a high-granularity
5D-tracking target, a first for this research program. The use of agentic-AI
will significantly improve the detector design feedback loop, optimizing PIONEER's sensitivity to rare decays.

To enable these approaches at scale, AI-based design and optimization tools are critical for both the scale and complexity of the challenges. Technologies such as differentiable simulations and surrogate models can enable gradient-based optimization, while active learning approaches like reinforcement learning and Bayesian optimization can help drive broader exploration of design spaces. To orchestrate such a design loop, agentic AI systems can be used to develop automated workflows that coordinate simulation, optimization, and analysis across multiple subsystems. Traditionally, these challenges have been addressed by disparate communities with disconnected tools, which limits the possibility of holistic optimization. Agentic systems can streamline and simplify complex design and simulation choices, manage iterative optimization cycles, and rapidly evaluate alternative configurations. By reducing manual intervention and enabling adaptive decision making, agentic AI can significantly accelerate the identification of optimal designs in large and complex parameter spaces.

In addition to software-driven optimization, AI-native experimental design naturally extends into robotics, automated laboratories, and advanced manufacturing, creating new opportunities to streamline the construction, testing, and validation of experimental hardware. Robotic assembly and automated quality assurance can significantly accelerate the production of detector modules, while AI-controlled test stands and smart laboratories enable rapid, closed-loop iteration between design, fabrication, and performance evaluation. Advanced manufacturing techniques, including additive manufacturing and precision automated machining, can further enable novel detector geometries and materials that would be impractical with traditional fabrication workflows. Integrating these capabilities into the AI-driven design loop closes the gap between conceptual optimization and physical realization, reducing development time, improving reproducibility, and lowering overall technical risk.

Beyond technical performance, large particle physics facilities operate at the multi-billion-dollar scale, making accurate and reliable cost estimation an essential component of the design process. AI-based large language models provide a powerful opportunity to build robust cost models by integrating empirical scaling laws, detailed engineering estimates, and data from past and ongoing large-scale projects. In addition, these tools can generate more detailed technical design reports and engineering plans with substantially less effort, improving cost realism early in the design cycle and helping to mitigate cost escalation over the lifetime of a project. By synthesizing heterogeneous sources of cost and schedule information, such models can support more reliable projections of total cost, construction timelines, and technical risk, and enable rapid evaluation of design tradeoffs from both performance and cost perspectives.

AI-native experimental design can reduce the technical and fiscal challenges of bringing these projects to life, and introduce new scientific capabilities. Developing common AI-enabled tools for performance optimization, automated workflows, detector design, and cost modeling would dramatically accelerate design iteration, reduce technical risk, and shorten the path from concept to construction.

\subsection{Grand Challenge 2: Intelligent Sensing and Instrumentation} 
%(1 page)}
\label{sec:sensing}
%[Nhan, Sarah, Giordon, Isobel, Ian; Perhaps add "data" to title?]
%
% Multi-messenger sentences [Ken]
%[Reviewers: Viviana]

Next-generation particle physics experiments are increasingly limited by how rapidly and intelligently data can be accessed, filtered, interpreted, and acted upon. AI-driven rapid data access is therefore becoming a critical enabler across the full chain of data acquisition, triggering, reconstruction, and analysis, with direct impact on physics sensitivity, discovery potential, and operational efficiency.

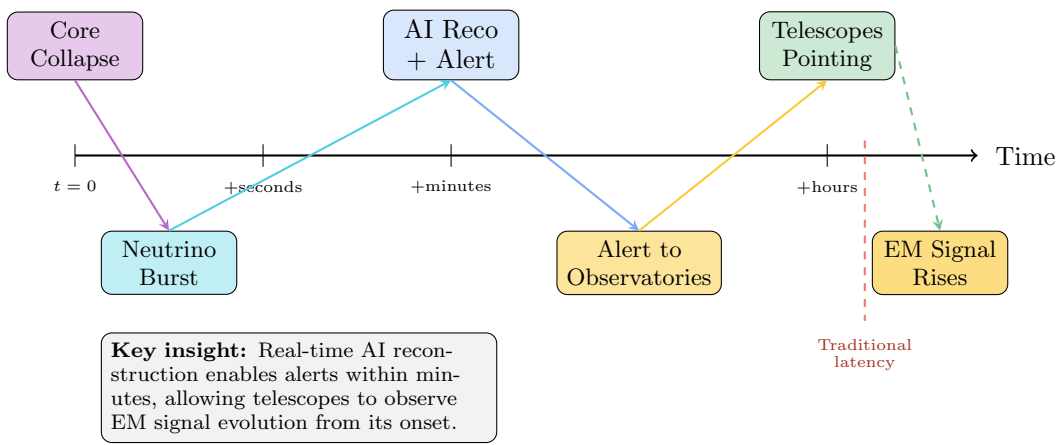
\begin{figure}[!htbp]
\centering

% === Color definitions (configurable) ===
\definecolor{aicolor}{RGB}{66, 133, 244}       % Blue for AI components
\definecolor{discardcolor}{RGB}{234, 67, 53}   % Red for discarded data
\definecolor{keepcolor}{RGB}{52, 168, 83}      % Green for kept data
\definecolor{alertcolor}{RGB}{251, 188, 5}     % Orange/yellow for alerts
\definecolor{neutrinocolor}{RGB}{0, 188, 212}  % Cyan for neutrino events
\definecolor{astrocolor}{RGB}{156, 39, 176}    % Purple for astrophysical events

% === Common styles ===
\tikzset{
    box/.style={
        rectangle, draw, rounded corners, 
        minimum width=2cm, minimum height=0.8cm, 
        align=center, font=\small
    },
    smallbox/.style={
        rectangle, draw, rounded corners, 
        minimum width=1.8cm, minimum height=0.7cm, 
        align=center, font=\footnotesize
    },
    aibox/.style={
        rectangle, draw, rounded corners, 
        minimum width=1.8cm, minimum height=0.8cm, 
        align=center, font=\small, fill=aicolor!20
    },
    discardbox/.style={
        rectangle, draw, rounded corners, 
        minimum width=1.8cm, minimum height=0.7cm, 
        align=center, font=\footnotesize, fill=discardcolor!20
    },
    keepbox/.style={
        rectangle, draw, rounded corners, 
        minimum width=1.8cm, minimum height=0.7cm, 
        align=center, font=\footnotesize, fill=keepcolor!20
    },
    alertbox/.style={
        rectangle, draw, rounded corners, 
        minimum width=1.8cm, minimum height=0.7cm, 
        align=center, font=\footnotesize, fill=alertcolor!40
    },
    arrow/.style={->, >=stealth, thick},
    darrow/.style={->, >=stealth, thick, dashed},
    title/.style={font=\bfseries\small},
    label/.style={font=\footnotesize\itshape},
}

% === Panel (a): Traditional vs AI-Native Data Acquisition ===
\begin{subfigure}[b]{\textwidth}
\centering
\begin{tikzpicture}[node distance=0.75cm and 1.5cm]

% --- Anchor point for the figure ---
\coordinate (origin) at (0,0);

% --- Traditional row (top) ---
\node[label, anchor=east] (trad-label) at (origin) {Traditional:};
\node[box, right=0.5cm of trad-label.east, anchor=west] (det1) {Detector};
\node[smallbox, right=0.75cm of det1] (trig1) {Hardware\\Trigger};
\node[discardbox, above right=0.75cm and 1.2cm of trig1.east, anchor=west] (disc1) {Discard\\$\sim$99.99\%};
\node[keepbox, below right=0.75cm and 1.2cm of trig1.east, anchor=west] (keep1) {Keep\\$\sim$0.01\%};
\node[smallbox, right=of keep1] (store1) {Storage};

% Traditional arrows
\draw[arrow] (det1.east) -- (trig1.west);
\draw[arrow] (trig1.east) -- (disc1.west);
\draw[arrow] (trig1.east) -- (keep1.west);
\draw[arrow] (keep1.east) -- (store1.west);

% X through discard box
\draw[discardcolor, very thick] (disc1.north west) -- (disc1.south east);
\draw[discardcolor, very thick] (disc1.south west) -- (disc1.north east);

% --- AI-Native row (bottom) ---
\node[label, anchor=east, below=2.5cm of trad-label] (ai-label) {AI-Native:};
\node[box, right=0.5cm of ai-label.east, anchor=west] (det2) {Detector};
\node[aibox, right=0.75cm of det2] (smart) {Smart Sensors\\ML on Edge Hardware};
\node[aibox, below=0.75cm of smart] (aiproc) {AI Processing\\+ Compression};

% Output nodes anchored to aiproc
\node[keepbox, above right=1.5cm and 1.5cm of aiproc.east, anchor=west] (pri1) {Priority\\Data};
\node[keepbox, right=1.5cm of aiproc.east, anchor=west] (pri2) {Compressed\\Stream};
\node[alertbox, below right=1.5cm and 1.5cm of aiproc.east, anchor=west] (alert) {Real-time\\Alerts};

% Final destinations anchored to outputs
\node[smallbox, above right=0cm and 0.8cm of pri2.east, anchor=west] (store2) {Tiered\\Storage};
\node[alertbox, right=0.8cm of alert.east, anchor=west] (ext) {External\\Observatories};

% AI-Native arrows
\draw[arrow] (det2.east) -- (smart.west);
\draw[arrow] (smart.south) -- (aiproc.north);
\draw[arrow] (aiproc.east) -- ++(0.75,0) |- (pri1.west);
\draw[arrow] (aiproc.east) -- (pri2.west);
\draw[arrow] (aiproc.east) -- ++(0.75,0) |- (alert.west);
\draw[arrow] (pri1.east) -- ++(0.5,0) |- (store2.west);
\draw[arrow] (pri2.east) -- (store2.west);
\draw[arrow] (alert.east) -- (ext.west);

% --- Annotation box anchored to bottom of figure ---
\node[draw, rounded corners, fill=gray!10, font=\scriptsize, align=center,
      below right=0.8cm and -2.5cm of aiproc.south, anchor=north] (annotation) 
      {Intelligence moves upstream; substantially less irreversible discard};

% --- Title anchored to top of figure ---
\node[title, above right=1.0cm and 0.75cm of det1.north west, anchor=south west] (panel-title) 
      {Traditional vs.\ AI-Native Data Acquisition};

\end{tikzpicture}
\caption{Comparison of traditional trigger-based data acquisition, which irreversibly discards the vast majority of data at the hardware level, versus AI-native continuous readout with intelligent on-detector processing, compression, and prioritization.}
\label{fig:gc2-acquisition}
\end{subfigure}

\vspace{0.8cm}

% === Panel (b): Multi-Messenger Alert Timeline ===
\begin{subfigure}[b]{\textwidth}
\centering
\begin{tikzpicture}[node distance=1cm and 1.5cm]

% --- Anchor point: left end of time axis ---
\coordinate (timeorigin) at (0,0);

% --- Time axis ---
\draw[->, thick] (timeorigin) -- ++(12,0) coordinate (timeend);
\node[font=\small, right=0.1cm of timeend, anchor=west] {Time};

% --- Time markers (relative to timeorigin) ---
\foreach \xshift/\timelabel [count=\i] in {0/{$t=0$}, 2.5/{+seconds}, 5/{+minutes}, 10/{+hours}} {
    \coordinate (tick\i) at ($(timeorigin)+(\xshift,0)$);
    \draw (tick\i) -- ++(0,0.15);
    \draw (tick\i) -- ++(0,-0.15);
    \node[font=\tiny, below=0.2cm of tick\i, anchor=north] {\timelabel};
}

% --- Event boxes anchored to time positions ---
% Row 1 (upper row of events)
\node[smallbox, fill=astrocolor!25, above=1cm of tick1, anchor=south] (sn) {Core\\Collapse};
\node[aibox, above=1cm of tick3, anchor=south] (aireco) {AI Reco\\+ Alert};
\node[smallbox, fill=keepcolor!25, above=1cm of tick4, anchor=south] (point) {Telescopes\\Pointing};

% Row 2 (lower row of events) - offset between ticks
\coordinate (between12) at ($(tick1)!0.5!(tick2)$);
\coordinate (between23) at ($(tick2)!0.6!(tick3)$);
\coordinate (between34) at ($(tick3)!0.5!(tick4)$);
\coordinate (after4) at ($(tick4)+(1.5,0)$);

\node[smallbox, fill=neutrinocolor!25, below=1cm of between12, anchor=north] (nu) {Neutrino\\Burst};
\node[alertbox, below=1cm of between34, anchor=north] (gwalert) {Alert to\\Observatories};
\node[smallbox, fill=alertcolor!50, below=1cm of after4, anchor=north] (em) {EM Signal\\Rises};

% --- Connecting arrows using anchors ---
\draw[arrow, astrocolor!70] (sn.south) -- (nu.north);
\draw[arrow, neutrinocolor!70] (nu.north) -- (aireco.south);
\draw[arrow, aicolor!70] (aireco.south) -- (gwalert.north);
\draw[arrow, alertcolor!80] (gwalert.north) -- (point.south);
\draw[darrow, keepcolor!70] (point.east) -- (em.north);

% --- Traditional latency line ---
\coordinate (latency-top) at ($(tick4)+(0.5, 0.2)$);
\coordinate (latency-bot) at ($(tick4)+(0.5, -2.2)$);
\draw[dashed, discardcolor!70, thick] (latency-top) -- (latency-bot);
\node[font=\tiny, discardcolor!70!black, align=center, below=0.1cm of latency-bot, anchor=north] 
     {Traditional\\latency};

% --- Key box anchored relative to figure ---
\node[draw, rounded corners, fill=gray!10, font=\scriptsize, align=left, text width=5cm,
      below=0.5cm of nu.south west, anchor=north west] (keybox) {
    \textbf{Key insight:} Real-time AI reconstruction enables alerts within minutes, allowing telescopes to observe EM signal evolution from its onset.
};

% --- Title anchored to top of figure ---
\node[title, above right=0.3cm and 1.25cm of aireco.north, anchor=south] (panel-title) 
      {Multi-Messenger Alert Timeline};

\end{tikzpicture}
\caption{Multi-messenger astrophysics timeline illustrating how real-time AI reconstruction of neutrino events enables rapid alerts to external observatories. Telescopes can be positioned before electromagnetic counterparts begin to rise, capturing physics that would otherwise be missed due to traditional processing latencies.}
\label{fig:gc2-multimessenger}
\end{subfigure}

\caption{Intelligent sensing and data acquisition for next-generation HEP experiments.}
\label{fig:gc2-sensing}
\end{figure}

A transformative development enabled by AI is the emergence of trigger-less or trigger-free data acquisition architectures, in which detectors operate in continuous readout mode and essentially all data are collected. In this paradigm, traditional hard trigger decisions, in which the majority of data are irreversibly discarded based on simple selection criteria, are replaced by real-time, AI-driven reconstruction and inference. Machine-learning models operating close to the detector can perform fast pattern recognition, timing reconstruction, and physics-aware feature extraction, enabling intelligent prioritization, compression, and routing of data without prematurely rejecting events classified as “background.” This fundamentally changes how experiments balance bandwidth, storage, and physics sensitivity, allowing rare, unexpected, or poorly modeled signals to be retained and studied. Realization of this goal will require developing flexible neural networks deployable in on-detector front-end and off-detector electronics, moving intelligence upstream, with the additional constraint that on-detector electronics typically needs to be radiation hard.

These approaches build on advances in smart sensors and electronics, including smart pixels and ML-assisted front-end systems (ML-on-Edge Hardware), which embed inference directly into the data acquisition chain. Combined with AI-guided reconstruction and high-dimensional, physics-aware compression, they enable dynamically and intelligently adaptive readout paths that optimize resource usage while preserving scientifically valuable information. Rather than throwing away data at trigger level, AI models learn compact representations that retain subtle correlations and rare features, enabling both targeted physics measurements and anomaly detection beyond predefined signatures.

%\begin{sloppypar}
%\end{sloppypar}
Trigger-less, AI-driven architectures are especially impactful for time-critical and data-intensive domains such as fast timing, neutrino physics, and multi-messenger astronomy. 
In the detection of transient astrophysical events—such as supernova neutrinos—real-time AI reconstruction enables rapid identification and characterization of signals and the immediate dissemination of alerts to external observatories, including optical telescopes, satellites, and gravitational-wave detectors such as LIGO. A rapid localization of the gravitational wave and/or neutrino signatures enables telescopes to be in position to search for the electromagnetic (EM) signature before it even begins to rise (the EM signal is typically delayed by a few hours due to the density of the surrounding material). Thus they can capture the electromagnetic signal's evolution from the very beginning, which contains a rich amount of physics and is often not captured. This capability is essential in constrained environments such as underground (e.g. DUNE) or remote (e.g. IceCube, LSST) experiments, where power, networking, and latency impose strict limits on on-site processing capabilities, data movement, and storage. AI tools could also speed up localization by routing data processing tasks away from sites experiencing computing cluster downtimes or poorly performing network routes. In this arena, every minute counts. Technologies for 
ML-on-Edge can be expanded to other domains that use remote sensing, such as geophysical distributed sensor monitoring networks such as SAGE~\cite{sage} and GAGE~\cite{gage}, biological and environmental monitoring networks such as NEON~\cite{neon} and satellites for observation, science, communication, and defense/security.
% https://www.earthscope.org
% https://www.unavco.org  GAGE
% https://www.iris.edu/hq/
% https://www.neonscience.org

Beyond immediate decision making, AI-enabled continuous readout makes it possible to selectively retain information that is currently impractical to store at scale, such as raw detector waveforms relevant for dark matter searches, solar neutrino measurements, neutrinoless double-beta decay searches, or core-collapse supernovae. By shifting intelligence from rigid trigger logic to adaptive, learning-based systems, experiments gain access to new classes of observables and entirely novel analysis strategies, while simultaneously improving data-taking efficiency, detector uptime, and long-term scientific return.

Finally, the incorporation of interpretability, uncertainty quantification, and robustness into real-time AI models is a critical emerging direction. As AI systems increasingly determine what data are compressed, prioritized, or permanently stored, transparent and explainable decision making becomes essential for validation, trust, and physics insight. Together, trigger-less architectures and AI-enabled rapid data access represent a paradigm shift in how particle physics experiments operate, enabling faster discovery, greater sensitivity to the unexpected, and more efficient use of large-scale detector and computing infrastructures.

%%%\newpage
%\subsection{Grand Challenge 3: Automated facility operations, data quality and calibration [Alternate: Self‑Driving Experiments: AI‑Enabled Operations \& Calibration(1 page)}
%\subsection{Grand Challenge 3: Self‑Driving Experiments: AI‑Enabled Operations \& Calibration}
%\subsection{Grand Challenge 3: Autonomous Experiments: Agentic Operations \& Calibration}
\subsection{Grand Challenge 3: Autonomous Experiments}
\label{sec:GC3}
%(1 page)}

%[Reviewers: Walter, Ken, Giordon, Terao]

This Grand Challenge targets using AI to improve and automate detector and facility operations, computing operations and calibration derivation that together consume substantial human effort through complex tasks and 24/7 expert support. High-quality, stable data is a prerequisite for physics analysis, and no downstream AI can recover information lost to poor data quality or downtime. Operational excellence underpins all other AI applications: even small gains in uptime, efficiency, or calibration speed translate directly into physics and discovery impact.

\begin{figure}[!htbp]
    \centering
    \includegraphics[width=0.48\linewidth]{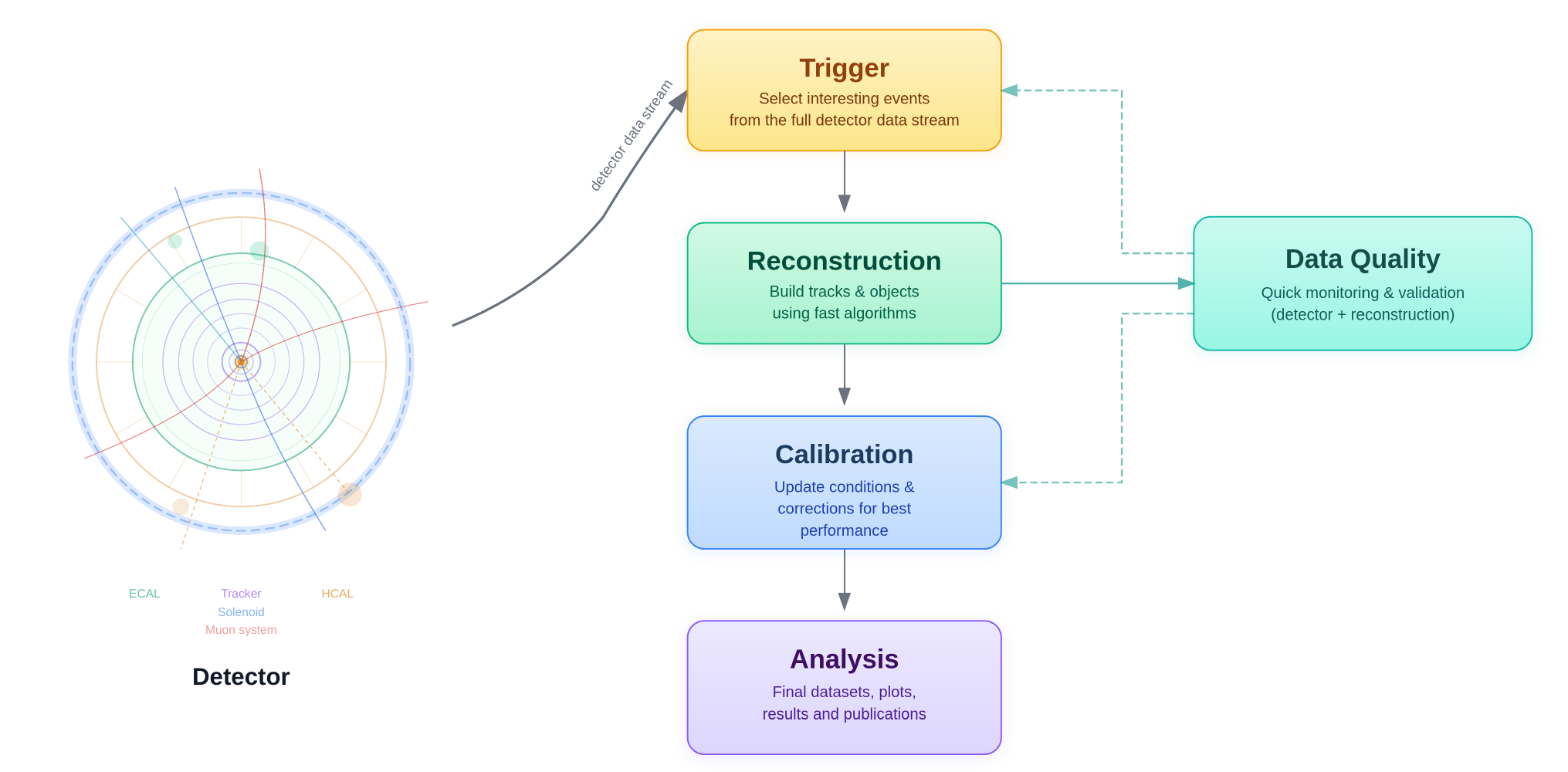}
    \includegraphics[width=0.48\linewidth]{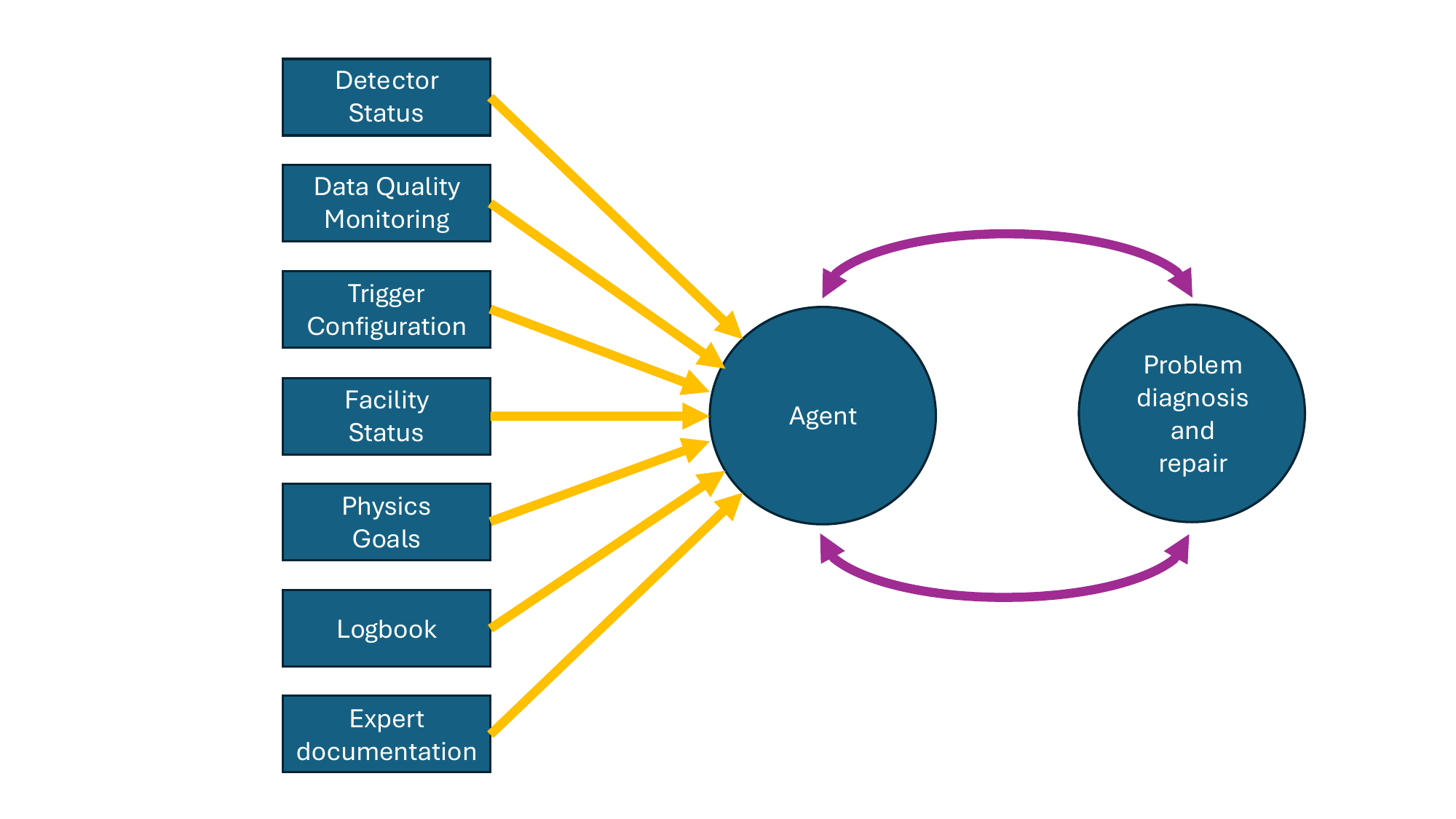}
    \caption{Data-taking, data quality and calibration flow diagram and how online and offline information might be used to reduce operational need, expert callbacks and calibration robustness.}
    \label{fig:gc3-fig}
\end{figure}

This vision aims to enable routine experiment operation with control-room staffing limited to detector and facility safety and decreasing reliance on dedicated expert interventions by roughly an order of magnitude. Distributed computing operations, data quality monitoring and calibration would be handled by automated systems, increasing data throughput and reducing the time to derive calibrations from months to days. Together, these advances could deliver up to 50\% less downtime, $10\times$ faster calibration cycles, and improved data quality through AI-driven monitoring, diagnosis, and operational decision support—while reducing the required personnel effort for calibration and data quality tasks by a factor of ten.

Today’s operations are personpower-intensive and fragile. Large detectors rely on hundreds of experts for 24/7 monitoring and triage. Each hour of detector downtime can cost on the order of $\sim$ \$300k in lost physics opportunity and increases the chance of missing a once-a-century/millennium event. Maintaining round-the-clock operations places substantial strain on personnel: shift work drives fatigue, errors, and steep training demands, increasing reliance on scarce expert support. Critical know-how is concentrated in a few individuals, and documentation is fragmented and often outdated. Data quality relies on teams manually reviewing thousands of fast-updating histograms with limited automation, while calibration can require 100+ people and takes months to years of effort, delaying physics analyses. Problems are often discovered only months after data taking, when recovery is impossible, and the effort required to recalibrate limits how often datasets can be improved.

AI will directly reduce downtime and labor by shifting operations from reactive to proactive. Anomaly detection can flag problems before they impact data, while operations assistants based on Large Language Models (LLMs) can combine real-time detector or beamline information with logbooks, log files and documentation to guide shifters based on previous events, shorten training, and reduce expert callouts. Predictive maintenance can forecast component failures hours to days ahead, enabling planned interventions instead of emergency downtimes.

AI will also modernize data quality and calibration. Automated Quality Assurance and control (QA/QC) can flag outliers for rapid human review, and real-time physics-level monitoring can provide immediate feedback on data usability rather than delays of hours to weeks. Agentic calibration workflows could shrink week-long campaigns to hours, improving consistency and reducing systematic uncertainties.
Similarly, AI‑driven assistants and agentic systems can play a transformative role by supporting distributed computing operations, including automated workflow diagnosis, intelligent resource utilization, user support, and real‑time monitoring across heterogeneous infrastructures. 

%
% Commented-out by Kazu, replaced by the paragraph below which added an impact on training a new generation + public outreach. Opportunities "to learn about detectors/accelerators" is precious but often limited to those who participated in design/production/installation/commissioning. Students who missed those only can learn from reading or talking to very busy experts w/ little time. 24/7 interactive agents can address that bottleneck.
%
%These tools are also a way to preserve institutional knowledge as senior experts retire: AI systems can be trained to capture procedures, diagnostics, and lessons learned, providing near-term gains in current experiments while strengthening the foundation for all downstream AI-driven physics.

% Kazu: Below is a paragraph w/ added texts
These tools also play a critical role in both preserving institutional knowledge as senior experts retire and training future generations of physicists at scale. AI systems can be trained to capture procedures, diagnostics, and lessons learned, providing near-term gains in current experiments while strengthening the foundation for all downstream AI-driven physics. These tools could offer 24/7 interactive access and deep scientific knowledge about facilities.

%which can effectively help to train researchers about critical physics knowledge behind. They may be fine-tuned and used to make learning about world-class scientific facilities more accessible to the broader particle physics community, including students, and even for the general public. 

%Revised accelerator/facilities paragraph from Heather
Parallel efforts are underway for accelerators and beamline facilities~\cite{HEP_ABP_Roadmap_2023}, closely related to those described above for detectors. Even modest improvements in beam intensity, stability, or quality, alongside reductions in unplanned downtime, translate directly into enhanced physics reach, and would increase both the quantity and quality of data available for analysis. The anticipated benefits mirror those foreseen for particle detectors: reduced operational labor demands, predictive maintenance, and improved knowledge preservation and training.

Key required capabilities include integrated multi-modal time-series analysis across subsystems, reliable root-cause inference for coupled failures, end-to-end automated pipelines, and uncertainty-aware anomaly detection that avoids alert fatigue. LLM assistants must be grounded in detector and facility operations to prevent hallucinations and will require training on internal experiment or accelerator information.

Near-term progress should focus on deployable pilots: cross-experiment operations AI working groups, subsystem-level automated calibration demonstrators, automation of workflow triage in computing operations and logbook-integrated shifter assistants. By HL-LHC commissioning and first beam at DUNE, experiments could target $\sim$30\% downtime reductions, continuous automated calibration, and AI-supported training. The long-term goal is supervised autonomy—experiments designed from day one around AI-assisted operations so humans spend less time responding to problem reports and more time doing physics.

%merged above
%Finally, beyond detector operations and calibration, autonomous experiments increasingly depend on complex, distributed computing and software ecosystems that span facilities, workflows, and user communities. AI‑driven assistants and agentic systems can play a transformative role by supporting distributed computing operations, including automated workflow diagnosis, intelligent resource utilization, user support, and real‑time monitoring across heterogeneous infrastructures. 

%seems like an entirely different challenge
%AI‑assisted code generation, refactoring, and validation can further accelerate the evolution of experiment software frameworks, reduce barriers to adoption of new computing models, and improve long‑term sustainability. Together, these capabilities extend the concept of autonomy beyond the detector itself, enabling experiments to operate as resilient, self‑optimizing systems that scale scientific productivity while reducing operational burden.

%%%\newpage

%\subsection{Grand Challenge 4: Automated and optimizable data analysis [Alternate: Accelerate Analysis for Discovery ?, Alternate 2: Accelerate Discovery, Alternate 3: Automated and Adaptable Data Analysis to Accelerate Discovery] (1 page)}
\subsection{Grand Challenge 4: From Data to Discovery}
\label{sec:GC4}

This grand challenge seeks to use AI to enable a paradigm-shifting 
breakthrough in particle physics through innovative data processing and analysis workflows. 
%not merely by making existing workflows 
%faster, but by making them systematically more optimal and scalable. 
Today, a single high-energy physics (HEP) analysis typically requires several 
{\em years} of sustained human effort dedicated to iterative development, tuning, 
validation and review. We aim to reduce the time required to 
design, refine, and evaluate an analysis by factors of 100–1000. These 
approaches will enable the community to explore a scientific landscape that is 
itself orders of magnitude larger. Instead of testing a few
hand-chosen benchmark scenarios, analyses could efficiently probe entire spaces of 
theoretical possibilities, mapping the behavior, capabilities, and limitations 
of whole families of models in a controlled and reproducible way. The proposed 
methods shift effort away from low-level technical mechanics toward 
formulating scientific questions, guiding optimization, and interpreting 
results—fundamentally transforming how HEP conducts science and extending its 
reach.

\begin{figure}[!htbp]
    \centering
    \includegraphics[width=0.8\linewidth]{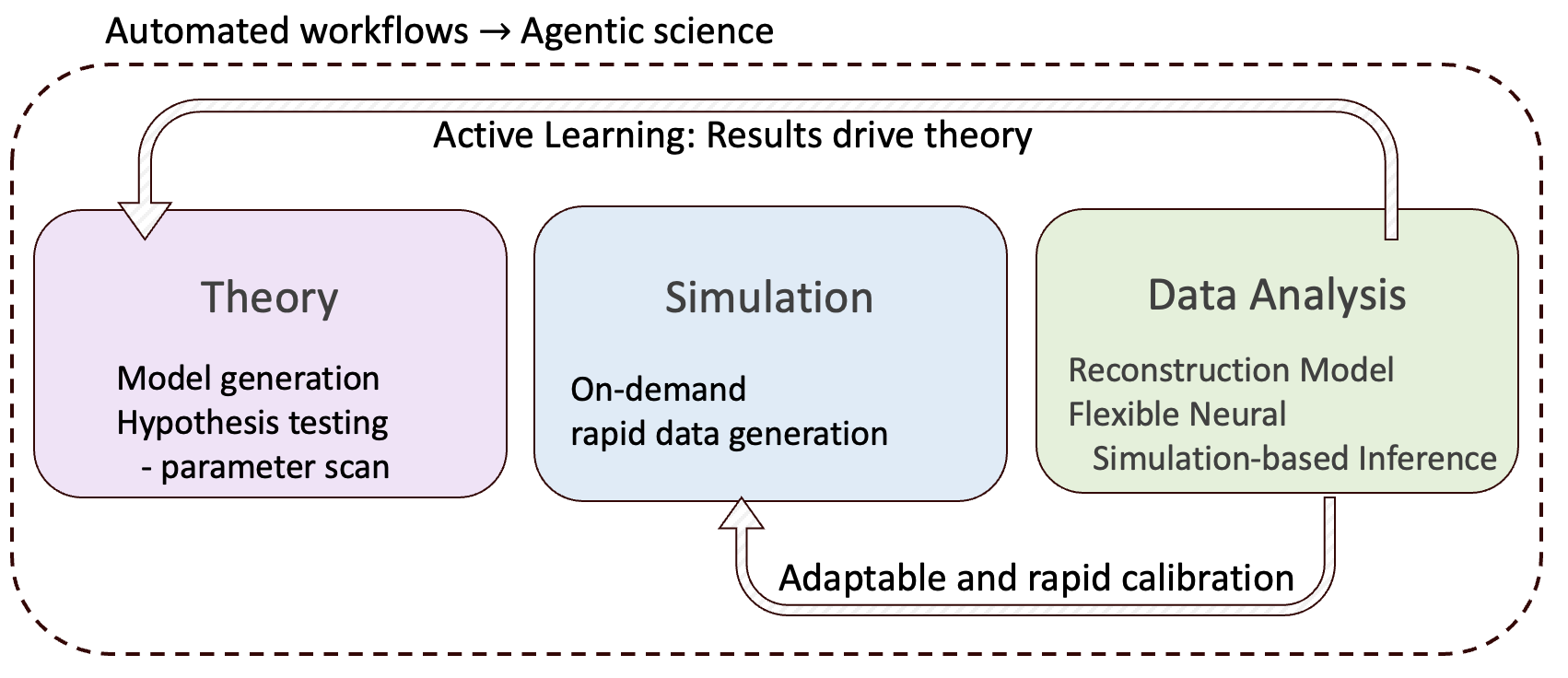}
    \caption{Automatic and optimizable data analysis diagram, adapted from~\cite{kaganheinrichtalk}}
    \label{fig:GC4_diagram}
\end{figure}

Realizing this new paradigm requires building systems capable of rapid, 
optimal, and automated searches and precision measurements across large 
datasets and extensive theory spaces, following the approach in 
Figure~\ref{fig:GC4_diagram}. AI tools can adapt the full analysis pipeline to 
a given physics objective, optimizing each stage to maximize sensitivity. The 
{\bf optimization} and {\bf automation} are the two pillars of this grand challenge, and 
can proceed independently: advances in one improve the performance of the 
other. Crucially, these techniques apply broadly, not just to one analysis.

%The first step in the analysis chain is theory selection, which can be automated either by constructing models directly from Lagrangians describing quantum field theories, using tools such as FeynRules, or by systematically probing existing models and exploring alternative hypotheses through parameter scans.

\paragraph{Transformational capabilities enabled by multimodal foundation models}

Multimodal foundation models offer a powerful way to optimize and 
coordinate the full physics analysis chain, enabling capabilities that are 
impractical with today’s task-specific models.
Today's
%the current practice of developing 
bespoke, narrowly trained models can be replaced with fine-tuning general-purpose 
multimodal foundation models for analysis-specific objectives. By pretraining 
on diverse data — including simulation as well as current and
archived experimental data — these models can be adapted to new tasks with 
dramatically reduced labeling requirements and substantially shorter 
development cycles. A defining feature of modern foundation models is their 
ability to transfer learned representations across interaction channels, event 
topologies, detector conditions, and even detector geometries. This 
transferability is further enhanced by their multimodal nature, 
allowing joint reasoning over heterogeneous inputs such as images, 
waveforms, point clouds, and structured physics features.

To fully realize these capabilities, adaptable foundation models must be 
developed at multiple levels of the analysis stack. At the reconstruction 
level, they process raw sensor data to enable robust and 
reusable physics object reconstruction. At the population level, shared 
representations improve signal discrimination and background 
characterization across analyses. At the inference level, foundation models 
can underpin simulation-based inference, unfolding, parameter estimation, and 
anomaly detection. This requires dedicated supervised and 
self-supervised training strategies across all data
abstractions of the data—from sensor-level measurements, to reconstructed 
particles, to full events—while preserving physical interpretability and 
uncertainty quantification.

\paragraph{AI-Accelerated Simulation for Analysis}

Simulation is central to nearly all particle physics analyses as an
essential bridge between underlying theory, detector response, and observed 
data. Modern experiments demand simulated samples that are 
not only accurate, but also rapidly generated, adaptable, and closely 
calibrated to data. In this emerging paradigm, simulation becomes an on-demand 
analysis component, where both event generation and the calibration procedures and associated uncertainties can be tailored to specific signatures and 
updated as understanding evolves.

Today, simulation remains a major bottleneck. Currently, precision measurements and new
physics searches require billions of Monte Carlo events, consuming vast
computational resources and often limiting statistical reach or the evaluation 
of systematic uncertainties that depend on modeling assumptions. In the future, these needs are expected to grow significantly, reaching trillions for experiments at the HL-LHC. 
Calibrating simulation to data—essential for controlling systematics—can 
itself be slow and labor-intensive. AI methods can transform 
simulation from a static, pre-computed resource into a dynamic, adaptive element 
of the analysis loop.

Both algorithmic and hardware advances enable this 
transformation. Algorithmic techniques such as simulation-based 
inference, generative models, hybrid physics–ML surrogates, domain translation 
between simulation and data, differentiable simulation, and systematics-aware
fast simulation provide orders-of-magnitude speedups while preserving physical
fidelity. These methods support on-demand generation of events and
enable rapid exploration of parameter and uncertainty spaces that are
impractical with traditional Monte Carlo workflows. In parallel, hardware-aware
approaches—including GPU-accelerated physics simulation, GPU-native digital
twins, and tight integration with HPC facilities—make high-fidelity simulation
scalable to the data volumes and detector complexities of
next-generation experiments.

These capabilities are particularly critical for DUNE, the
HL-LHC, and future collider facilities, where 
simulated data needs far exceed the growth of conventional computing
resources. Importantly, many of these AI-enabled simulation approaches are
differentiable, allowing gradient-based optimization and inference to be
integrated directly into the analysis. Within agentic workflows,
such ``smart'' simulations enable active learning across the full analysis
chain—from event generation through detector response and
calibration—effectively treating simulation and inference as a
unified inverse problem. This shift promises dramatic efficiency gains
and adaptive, statistically powerful analysis strategies that are 
highly responsive to emerging physics insights.

\paragraph{AI-Accelerated Reconstruction for Analysis}

Reconstruction plays an important role in determining the physics reach of experimental
data. It transforms raw or simulated signals—hits, waveforms, images,
and timing information—into analysis-ready objects such as tracks, clusters,
vertices, and particle-identification features, shaping both sensitivity and
flexibility. For general-purpose detectors, this process is a
primary discovery lever across diverse physics signatures.

%Machine learning already plays a critical role in modern reconstruction. Learning-based algorithms have been widely adopted to improve the speed, accuracy, and robustness of reconstruction tasks, often providing calibrated uncertainty estimates alongside point predictions. Much of this progress has come from small, task-specific deep-learning models with well-understood inductive structure and performance characteristics, which now form reliable and widely deployed reconstruction components.

Building on the current foundation of task-based models and procedural reconstruction
algorithms, large multimodal models offer a natural extension by integrating and
coordinating these components within a broader reconstruction ecosystem. Such
models can be fine-tuned to jointly reason over heterogeneous sensor inputs and
the intermediate representations produced throughout the reconstruction chain,
learning shared representations of detector response and particle signatures
reusable across multiple downstream tasks.
Ultimately, this points toward foundation models trained on
extensive, heterogeneous collections of sensor-level data and reconstruction
artifacts. By learning general and transferable representations, these models
enable homogeneous pipelines that can be shared across
analyses or adapted to specific use cases with limited additional training. In addition, there are problems in reconstruction, such as tracking, vertexing, clustering, where procedural reconstruction models still obtain superior performance over task-based models. Improving on procedural methods using AI likely requires either larger models or more advanced AI techniques.

%Because these representations span detector subsystems—and potentially experiments—they naturally support rapid domain adaptation, for example from simulation to data or across detector upgrades, reducing duplicated model development, mitigating systematic uncertainties, and shortening the human-driven iteration cycle of analysis development.

Beyond incremental gains, AI also enables new paradigms
that transcend traditional stage-by-stage pipelines. Probabilistic
reconstruction 
%infers latent physical
%quantities from raw detector data, producing 
infers posterior
distributions of latent physics quantities from raw data that naturally propagate uncertainties and correlations through
the analysis chain. Related end-to-end or bypass approaches map sensor-level
data directly to higher-level physics quantities or likelihoods to reduce
information loss and better align reconstruction with physics objectives.
Complementing these ideas, standardized AI embeddings for reconstructed objects
and events provide compact, detector-aware representations that can 
serve as a common interface between reconstruction, simulation, and inference
across analyses and experiments.
Advances in computational scalability and deployment on
accelerators—such as GPUs, FPGAs, and ASICs—will further extend these
capabilities into low-latency trigger systems and streaming analyses. AI-based approaches could also link across different levels of detector resolution, granularity and bandwidth, for example connecting data from real-time processing at the trigger to precision reconstruction offline. Integration of agent-orchestrated workflows with fast AI-based simulation and
AI-accelerated reconstruction will
increase physics reach, flexibility, and discovery potential.

\paragraph{AI for Statistical Interpretation}

The final stage of an analysis is statistical interpretation,
where experimental observations are compared with theoretical predictions.
%Despite increasingly high-fidelity simulations, 
This step is often limited by
the difficulty of solving the inverse problem: realistic likelihoods are
frequently intractable, forcing traditional approaches to rely on
low-dimensional summary statistics that lose significant information.
Neural simulation-based inference (NSBI) addresses this by
learning the mapping between observables and model parameters directly from
simulation, constructing neural likelihoods, likelihood ratios, or posteriors that
%bypassing explicit likelihood evaluation to 
enable flexible,
information-rich inference for both discovery and precision measurements. 
Related advances are also transforming unfolding, which connects detector-level
observations to theory-level quantities at the distribution level,  into a modern, more robust and scalable inference task driven by generative models and simulation-based learning.
In parallel, anomaly-detection offer a complementary,
model-agnostic approach by operating directly on data without relying on
specific simulations or theoretical hypotheses. In addition, uncertainty quantification is essential to produce physics results and a major challenge to ensure that AI-based predictions are reliable and interpretable.

Realizing the full potential of these methods requires active learning and
closed-loop workflows. By using intermediate analysis results to guide theory
selection, simulation, and further inference, active learning—together with AI
agents—enables scalable automation across complex hypothesis spaces and closes
the loop between data, theory, and interpretation. 

%More broadly, AI can fundamentally 
These methods will also strengthen the interface between theory and
experiment, enabling rapid exploration of high-dimensional
theory spaces, including effective field theories and simplified models,
yielding dynamic interpretations that evolve as new data or theoretical ideas
emerge. These approaches can identify poorly constrained regions of parameter
space, expose tensions between datasets, and suggest targeted re-simulation or new
measurements.
%and potentially support hypothesis generation by uncovering latent patterns or
%correlations. 
Extending these capabilities across experiments enables global
interpretations of shared theoretical models and a far broader exploration of
particle physics than is feasible today

\paragraph{Agentic workflows}

Agentic AI automates the scientific workflow itself. Given a
high-level goal—such as searching for dark matter or measuring the Higgs
self-coupling—a multimodal, agent-based system built on large language models
can organize the end-to-end analysis: selecting parameter points to evaluate,
incorporating active learning to guide hypothesis exploration, and iteratively
refining the analysis strategy as results accumulate. The agent orchestrates the
technical tasks, including Monte Carlo generation, analysis configuration
and execution, systematic evaluations, and statistical modeling. The
physicist directs the scientific process by defining the analysis
goals, assumptions, and constraints that guide the automated system. This
allows advances in underlying capabilities—such
as faster simulation, improved foundation models, or more powerful inference
techniques—to appear immediately in the workflow without manual
re-engineering by the analyst. In addition, the use of large language models may offer a complementary interpretability layer that integrates intrinsic, model-level mechanisms with prompt-driven, human-in-the-loop interaction. Additional considerations when developing agentic workflows include the ability to verify the correctness and provide feedback, and the ability to generalize, providing reusable solutions that are often more efficient and less error prone.

%\paragraph{GC4 Summary:} 
\paragraph{Summary:} The result is an analysis chain that is not merely 
automated, but adaptive, interrogable through physicist prompts, optimizable 
at every stage, and scalable across the full breadth of particle physics
measurements and theory space.

\subsection{Cross-cutting Themes and Emerging Opportunities} 

{\bf Cross-cutting themes and areas of active community interest:} While many of the community’s contributions aligned naturally within the four Grand Challenges described above, some of the proposed projects either fall outside those categories and/or have recurring themes that cut across them. These merit explicit acknowledgment and are discussed here. First, neutrino interaction modeling and cross section inference emerged repeatedly, with proposals to combine simulation-based inference, generative surrogates, and domain translation to learn robust, data driven models; and to build public neutrino benchmarks and Open Data challenges that standardize tasks, metrics, and leaderboards. In parallel, inputs emphasized theory–experiment integration, notably Standard Model Effective Field Theory (SMEFT/EFT) inference pipelines and workflows aware of parton distribution functions (PDFs) that connect lattice Quantum Chromodynamics (QCD) and phenomenology to experiment via reusable latent representations and uncertainty-aware inference. A third cluster, not explicitly included in the Grand Challenges, centers on Monte Carlo including GPU-accelerated event generation. Complementing these are proposals to make data truly AI-ready, with curated, federated Open Data, common tokenized representations, and inference-as-a-service deployments that carry models across facilities; and to embed AI in operations and QA/QC through anomaly-aware data quality monitoring, agentic shift assistants, and vision based fabrication/assembly checks. Finally, several inputs highlight co-design of detectors and materials -- from smart pixels and embedded FPGAs to ML-guided discovery of scintillators and optical interfaces -- linking instrumentation innovation directly to AI-driven design loops. Together, these threads reinforce the need for solutions that span design, operations, analysis, and long-term stewardship of data and knowledge. The cross-cutting AI technologies and the cyberinfrastructure that underpin these projects and the four Grand Challenges are discussed further in Sections~\ref{sec:technical-foundations} and~\ref{sec:infrastructure}, respectively.

{\bf Interplay with theory and the case for coordination:} While our whitepaper concentrates on experimental particle physics, our community input underscores that AI-enabled discovery depends on a tight feedback loop between theorists and experimentalists: EFT and PDF advances inform generator developments and analysis objectives; differentiable and generative simulators shorten the path from theory hypothesis to experimental test; and curated, AI-ready Open Data lowers barriers for joint inference across datasets and facilities. The same inputs point to shared needs that exceed any single project: common data/representation standards and benchmarks; portable training and inference services; agentic workflow orchestration; and operations toolkits that reduce downtime and preserve institutional knowledge. These cross-cutting opportunities directly motivate the next section: a larger, coordinated collaboration that federates expertise across labs and universities, pools computing and data services, aligns with emerging national infrastructure, and scales workforce development so that each experiment benefits from shared tools, sustained support, and the ability to marshal greater collective effort on every project.

\newpage

\section{Collaboration - Building a National Effort}
\begin{figure}[!htbp]
    \centering
\includegraphics[width=1\textwidth]{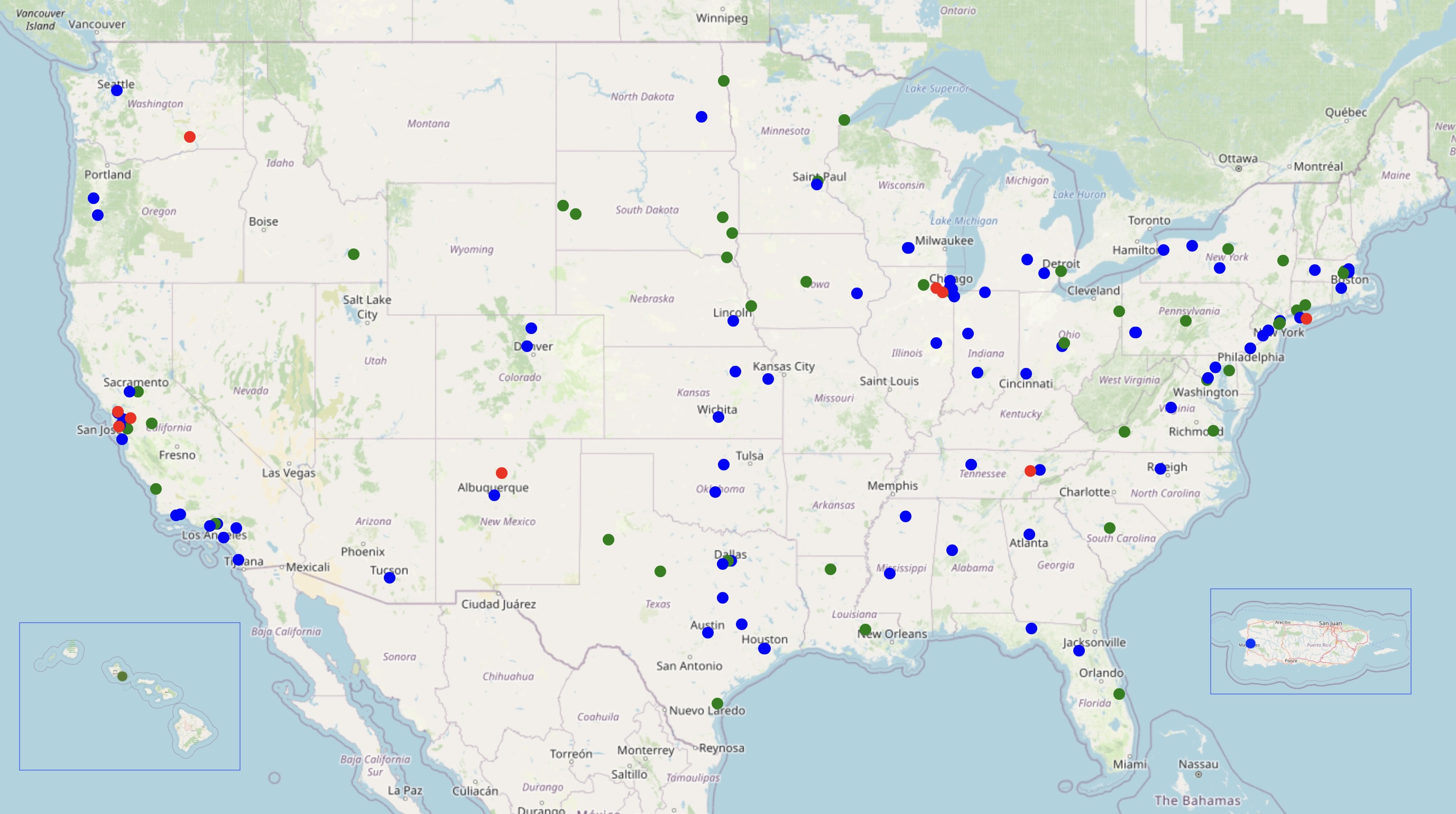}
    \caption{U.S. Institutions (blue and green dots = universities, red dots = DOE labs) that are currently part of large national-scale funded projects: U.S. ATLAS Operations (DOE/NSF, 33 institutions), U.S. CMS Operations (DOE/NSF, 44 institutions), DUNE Operations (DOE, 16 institutions), HEP-CCE (DOE, 4 labs), IRIS-HEP (NSF, 16 universities), A3D3 (NSF, 12 universities). Additional universities that are part of the U.S. ALICE, U.S. ATLAS, U.S. CMS and (U.S.) DUNE scientific collaborations are shown in green. Taken together, these overlapping experiments and projects tightly connect 9 DOE labs and 124 different universities and colleges.}
    \label{fig:collabmap}
\end{figure}

%(1 page)
\label{sec:collaboration}
%[Peter, Nhan, Sarah?]
\noindent
We envision creating a single \textbf{national-scale collaboration}
that brings together U.S. national laboratories, universities, and
diverse particle physics experiments to build a shared AI-native
research ecosystem for particle physics together with industry
partners. A national-scale collaboration is achievable. The U.S.
particle physics community already has extensive experience with
large-scale collaborations between U.S. universities and DOE national
labs as well as international partners, as shown in Figure~\ref{fig:collabmap}. For example, the U.S. CMS
collaboration includes two DOE labs (FNAL, LLNL) and 51 universities,
the U.S. ATLAS collaboration involves five DOE labs (BNL, ANL, LBNL, LLNL,
SLAC) and 42 universities and the U.S. ALICE collaboration involves two DOE labs (LBNL, ORNL) and 11 universities. Over the past three decades, these
collaborations, with international partners, have successfully built, operated, upgraded, and
delivered scientific results from the LHC experiments, and the DUNE-US organization is largely modeled after them.  This 
national-scale collaborative model enables significant resources to be
targeted at R\&D that moves quickly and strategically, guided by
the Grand Challenges, while also providing a fast and structured
path for universities to get involved.

%In the U.S. experimental particle physics ecosystem, national
%laboratories and universities play complementary roles. National
%laboratories provide large-scale production capabilities, access
%to advanced technological facilities, and highly skilled technical
%staff necessary for the realization of complex projects.
%Universities contribute in areas such as cutting-edge R\&D, workforce
%development, and the active involvement of students and postdoctoral
%researchers, enabling rapid innovation and sustained intellectual
%vitality. National laboratories can further serve as hubs that
%foster collaboration among geographically proximate universities,
%strengthening regional and national research networks. 
The greatest scientific impact is achieved when laboratories and 
universities are equal, tightly-integrated partners. 
%and tightly integrated via a common scientific
%goal to bring together academic innovation and training with
%laboratory-scale implementation and long-term stewardship.
By aligning academic innovation and training with laboratory-scale
implementation and stewardship, the HEP community can build a national 
research network and ensure that cutting-edge R\&D translates into the 
long-term impact required for the field’s Grand Challenges.

\paragraph{How would a national-scale collaboration work?}
The success of this collaboration requires more than the typical
institution-by-institution ``bring your own budget'' model used by
many traditional scientific collaborations.  
While bottom-up participation will remain essential,
it is not sufficient on its own to support the scale, coherence
and sustained effort required here. Thus we propose a \textbf{jointly managed} 
core effort across multiple large funding partners—DOE (Genesis
Mission), NSF, and private foundations—should be pursued. 
Modeled on the successful U.S. LHC operations programs, each of which 
manages O(\$40M/year) from two funding partners (DOE and NSF), this
structure supports the definition, pursuit, and evolution of coordinated 
``Grand Challenge'' activities, developed in close partnership with the 
experiments. DOE national laboratories provide scale, advanced computing
resources, and long-term stewardship of complex projects. Universities drive
innovative R\&D and workforce development through deep engagement of students and
postdoctoral researchers, delivering complementary strengths in rapid
prototyping, cross-disciplinary experimentation, and community training. NSF’s
mission and community enable expansion beyond particle physics and provide
pathways to engage other domain sciences, translating DOE investments more
broadly across the U.S. research and education ecosystem. Strategic partnerships
with industry ensure rapid access to state-of-the-art AI technologies and best
practices, enabling efficient co-design and technology transfer. Targeted
investments by private philanthropic partners can further support innovation,
education, and translation. 
In the example model leadership is drawn from both
labs and universities, and an annual planning and budget process
(spanning the DOE and NSF streams) 
%for U.S. ATLAS \& U.S. CMS) 
updates funding for collaborating
institutions based on progress towards milestones and deliverables
guided by the Grand Challenges.  In the case of the national-scale
AI collaboration, an additional challenge is delivering to multiple
experiments with different scientific objectives and timelines.
This will be informed by, and built on, the (cross-experiment)
experience from other community national-scale R\&D projects such
as the DOE-funded HEP-CCE~\cite{hep-cce} project,
NSF-funded Institute for Research and Innovation in Software for High Energy Physics (IRIS-HEP)~\cite{iris-hep,iris-hep-nsf-award,iris-hep-nsf-renewal}
and A3D3~\cite{a3d3-web, a3d3-nsf-award} institutes. 
%and DOE ASCR supported
%projects whose goal is to accelerate HEP discoveries through
%high-performance computing, such as the Scientific Discovery through
%Advanced Computing (SciDAC) program~\cite{SciDAC} and HPC and network
%facilities, such as NERSC that provide significant computing
%allocations to particle physics experiments.  
Experience with the
creation of national or regional physical hubs to support distributed
collaboration will be drawn from the Fermilab LHC Physics Center,
the U.S. ATLAS Analysis Center (at LBNL, BNL and ANL) and the NSF-funded 
IAIFI AI Institute~\cite{iaifi}. Leadership from all of
these efforts have been involved in this whitepaper.

\paragraph{What scale of collaboration is needed?}
Achieving the ambitious goals of a national AI collaboration
in particle physics will require substantial scale. Guided by
the structures of the U.S. high-energy physics program,
we estimate that a vibrant national program requires
a core effort 
%(analogous to the operations programs) 
of at 
least 120 full-time equivalents over 5 years, with roughly
twice that number of contributors.  A core effort of this size
or larger, involving both universities and DOE labs, aligns with the
scale of existing U.S. collaborations and projects. 
This is an estimated scale, reflecting the scope and ambition of 
the Grand Challenges, rather than a detailed bottom-up effort calculation. 
While the effort will include research components, these activities will be
explicitly guided by the Grand Challenges with the aim of building
an AI-native research ecosystem supporting the full lifecycle of
particle-physics experiments.  At this scale, the collaboration
core would also serve as an intellectual hub for many 
smaller separately funded R\&D efforts from across the broader 
community as well as providing a clear point of contact for 
international and industry partnerships. Such an AI collaboration
is designed to augment, not replace, the existing experimental operations programs.

\begin{comment}
%Peter what do you think about adding this kind of thing?
\paragraph{Collaboration initiatives:}
Several initiatives to build the collaboration include workshops to disseminate results and plan activities, Hackathons where groups come together to work on specific topics, data challenges which are virtual events where several solutions for topics can be explored, tutorials and hands-on training that provide direct experiences on which more advanced expertise can be built, and finally schools which are often broader and aimed at early stage researchers and students. These initiatives are connected to workforce development efforts, as described in Section~\ref{sec:workforce}.
\end{comment}

%XXX Clarify that this
%doesn't replace experiment specific integration and operations. Not
%bottom-up FTE calculation. XXX

%\paragraph{Collaboration Summary:} The HEP community’s historical head start in data-
%intensive science, combined with its mature organizational structures and diverse range 
%of experimentation, positions it uniquely to create such a national-scale effort. 
%The collaboration will leverage these strengths to develop new methodologies for 
%data-enabled discovery, ensuring the sustainability and growth of the field 
%while contributing broadly to the physical sciences and to the creation of an AI-enabled workforce.

\paragraph{Collaboration Summary:} The experimental particle physics
community has built, over many decades, organizational structures
linking every major U.S. research university, many undergraduate institutions
and nearly all DOE Office of Science labs.
%connecting essentially every major U.S. research university as
%well as many undergraduate institutions and nearly all of the DOE Office of
%Science laboratories.  
This community’s historical head start
in data-intensive science, combined with these mature structures
and diverse range of experimentation, uniquely positions it to
create a national-scale effort.  The collaboration we describe will
leverage these strengths to develop new methodologies for data- and
AI-enabled discovery, while contributing broadly to the physical
sciences and an AI-enabled workforce.
%Indeed the Office of Science laboratories have their origins and ongoing strong ties 
%to the nuclear \& particle physics research and accelerator science.
%ensuring the sustainability and growth of the field 

\newpage

\section{Workforce Development}
\label{sec:workforce}
% ADVANTAGES FROM A NATIONAL COLLABORATION SPANNING LABS/UNIVERSITIES
% - students participate in a cutting edge AI research project
% - students can leverage cutting edge AI for their particle physics research
% - students gain from both lab and university mentors
% - larger context for traditional REU/SULI programs
% - center of gravity for providing students opportunities, including
%   as they progress in their careers
% - structure to provide a coherent training and shared standards for that training
% - students gain access to expertise (and visibilty!) across the entire 
%   collaboration, not only their university group or experiment
% Possible activities: 
% - summer student and/or senior project level AI-related research projects
% - development of course modules and curricula on cutting edge AI topics for 
%   use in multi-university settings 
% - 2 year traineeships for PhD students (analogous to DOE Traineeships)
% - summer schools and hackathons to develop skills further and build cohort experience

Training and workforce development is key for the success of this initiative. The national-scale collaboration we propose will be capable of providing a structured pipeline for students and early-career researchers at all levels, from undergraduate through post-doctoral stages. Unifying efforts across universities and labs via such a collaboration will also transform fragmented training into a centralized and coherent ecosystem that both meets the needs of the community and also contributes to building a national AI-enabled workforce. By scaling beyond individual university or lab-level efforts, a national workforce program institutionalizes training and skill transfer, draws talent from all parts of society and ensures the long-term sustainability and impact of the AI-enabled research ecosystem.

Partnerships between universities, DOE national laboratories, and industry will be formed around scientific discovery and technology development. On one hand, the engagement between universities and industry partners will allow universities to serve as think tanks for U.S. technology companies, ensuring leadership in the global market. In the other direction, close collaboration with industry partners will allow students and early career researchers to train with the most modern AI tools and technologies. Universities and national laboratories have a long tradition of close engagement in workforce development, particularly in particle physics. These ties will be further strengthened by involving students in advanced AI projects in the labs, ensuring that they graduate with first-hand experience in the development of new technologies.

%DOE can bring a many educational opportunities to engage in cutting edge research with experts in specialized areas. Massive expansion of support for programs such as the DOE SCGSR, the Fermilab guest and visitors program, etc can provide such avenues. Identifying ways to improve the preparation of students and early career researchers, and reduce the burden on scientists at the labs and in industry can help  

Particle physics is well suited for dual-competency training with several applications that serve as testbeds for learning and developing AI methods. A key feature of this area is the availability of large datasets which can be used for AI. Large-scale experiments in particle physics make the data available in efficient data formats distributed in several DOE laboratories and universities across the country. These datasets offer a controlled environment to learn and develop new ideas in AI/ML. 

%The  collaboration can establish consistent training, shared standards
%within the community, and clear 
%career pathways to bridge the gaps between academic coursework at
%the universities and a research environment involving the national 
%laboratories and industry. As described in Section~\ref{sec:collaboration},
%such a collaboration will have connections across U.S. academia.

Workforce development will naturally proceed in two overlapping and synergistic approaches. Participants will contribute to R\&D projects to build the AI-enabled research ecosystem, gaining hands-on  experience with foundational AI tools and our national infrastructure. As the projects mature, they will increasingly focus on using these capabilities to accelerate scientific discoveries and technology development. DOE laboratories and Universities should be equal partners in this training pipeline with each managing different aspects in a coordinated fashion. We envision impact in different ways at different levels of education and training.

\paragraph{Undergraduate education:} Undergraduates will develop core knowledge and interest in our domain science (particle physics measurement and discovery) and AI-literacy 
simultaneously. There are several courses on AI fundamentals already available that include topics in mathematics, statistics, and computer science. These courses could be offered as part of the preparation for physics students, supplementing the mathematical methods and computational physics curriculum. The foundational training acquired through coursework is supplemented by particle physics domain-specific training and practical experience that connects the more general courses to specific research and technology development programs in AI for physics applications. Given the breath of this training, we encourage physics students to pursue dual-competency degrees. Traditional undergraduate research programs such as NSF Research Experiences for Undergraduates (REU) and DOE Science Undergraduate Laboratory Internships (SULI), as well as HEP-specific programs such as the Program for Undergraduate Research Summer Experience (PURSUE) and Summer Undergraduate Program for Exceptional Researchers (SUPER) organized by the U.S. CMS and U.S. ATLAS operations programs, and the IRIS-HEP Fellows program, can be enhanced to bridge from academic coursework to a cutting-edge research environment (in both physics and AI tools) and provide a broader context and numerous AI-related projects beyond traditional detector development or data analysis. Undergrads will connect via summer programs and via (funded or unfunded) independent study and/or senior thesis activities. Post-baccalaureate and bridge programs help recent bachelor degree recipients to gain the research skills and academic credentials needed to become competitive applicants to graduate programs and industry. Opportunities such as the A3D3 post-baccalaureate program~\cite{a3d3-postbac} can be extended to a broader set of research programs and incorporate both industry and lab partners. 

\paragraph{Graduate education (Masters/PhD):} 
Graduate researchers will grow into scientists with dual physics/AI competencies through a formalized co-mentorship model that brings together university faculty, national laboratory staff, and strategic industry partners. Building on and scaling the successful two-year DOE Computational HEP Traineeship 
programs~\cite{tac-hep,watchep,c2-the-p2}, this effort will embed PhD students within large, multi-institutional teams working on the Grand Challenge problems. In addition, we envision targeted annual training modules—often difficult for individual universities or laboratories to offer—that will provide up-to-date, practical instruction for integrating AI across the research ecosystem. 
The DOE Science Graduate Student Research (SCGSR) program can be expanded to support graduate students to spend an extended period at the DOE national labs to pursue their research program and develop advanced skills. Programs supporting practicums at DOE national labs, such as the DOE Computational Science Graduate Fellowship, can also be expanded to include such practicums in industry partners. 

\paragraph{Postdoctoral education:} By leveraging their competencies in both particle physics and AI technologies, postdocs will be well positioned to play key roles as technical architects of elements of the research ecosystem. Participation in this project will allow them to build leadership skills and prepare for roles as future principal investigators or as senior technical leads in national labs or in industry. 

%\begin{itemize}
%\item DOE laboratories and Universities need to be equal partners, including funding flowing directly to groups of universities
%\end{itemize}

\paragraph{Potential Training Scale:} 

As part of the Genesis Mission, for example, the DOE has described a national goal of training 100,000 scientists and engineers over the next decade ``to lead the world in  AI-powered science, innovation, and  applications'', and is currently seeking input on workforce development to achieve this 
goal~\cite{DOE-RFI-workforce-development}. The national-scale collaborative project we describe—organized around 
Grand Challenges and spanning multiple DOE laboratories and universities 
across the U.S.—offers a unique and concrete opportunity to 
contribute significantly to this goal of developing the next generation of 
AI-literate scientists.
The United States has produced an average of 215 PhDs in 
particle physics each year~\cite{NSF25-349, ncses2021doctorate} over the 
past 15 years, with some annual variation as new experiments begin 
taking data. The number of undergraduate students potentially engaged in 
this research is likely an order of  magnitude larger. 
Over the next decade, this equates to roughly 2,000 PhD and 20,000 
undergraduate students that would benefit directly from this collaboration 
and its AI-enabled research ecosystem.
If closely related nuclear physics experiments and students were also
engaged in the collaboration, these numbers would be 50\% higher. 
This represents a potential significant
contribution to the national Genesis Mission objectives. 
By involving students and early-career researchers directly in the
creation and use of such a cutting-edge AI-native research ecosystem, 
we will simultaneously drive transformative discovery and help 
secure long-term U.S. leadership in AI-enabled fundamental science.

\paragraph{Building on existing efforts:} In addition to the existing
workforce development programs mentioned above, the particle 
physics community has been a leader in building national level 
programs due to its national scale scientific collaborations and
large projects. For example, the DUNE collaboration's 
workforce development program~\cite{dune-tech} has one thrust 
geared towards undergraduate students and another towards 
early-career researchers. A number of annual summer schools focus
on AI and related technology and help build a cohort experience
for participating PhD students and postdoctoral researchers. These 
include the annual IAIFI summer school~\cite{iaifi-summer-school}, the
Machine Learning for Fundamental Physics School (ML4FP) school~\cite{ml4fp-2025} and 
the Computational and Data Science Training for High Energy 
Physics (CoDaS-HEP) summer school~\cite{codas-hep}. Complementing these 
efforts, the long running QuarkNet program~\cite{quarknet} provides a nationally 
coordinated education and outreach program that engages high school 
teachers and students, connects classrooms to frontier particle 
physics research, and thus further strengthens the long-term STEM 
workforce pipeline.

\newpage

\section{AI Technologies for High Energy Physics}
%(2 pages)
\label{sec:technical-foundations}
% [Lead: Michelle, with Nhan?, Michael?, Kyle?, Lukas?, Kazu?, Hiro?, Paolo? Jianming]
% What AI we will use to solve the Grand Challenges\\
% \begin{itemize}
% \item{Foundation Models}
% \item{Agentic AI}
% \item{SBI}
% \item{Dimensionality reduction/Realtime systems --(NT: hardware codesign?)}
% \item{Differentiable modeling}
% \item{Self-learning/adaptive systems, active/adaptive learning}
% \end{itemize}
% (include foundation models and techniques here like end-to-end and adaptable analysis)

 %Techniques, references

 %\vspace{0.5cm}

\begin{figure}[!htbp]
    \centering
    \includegraphics[width=0.90\linewidth]{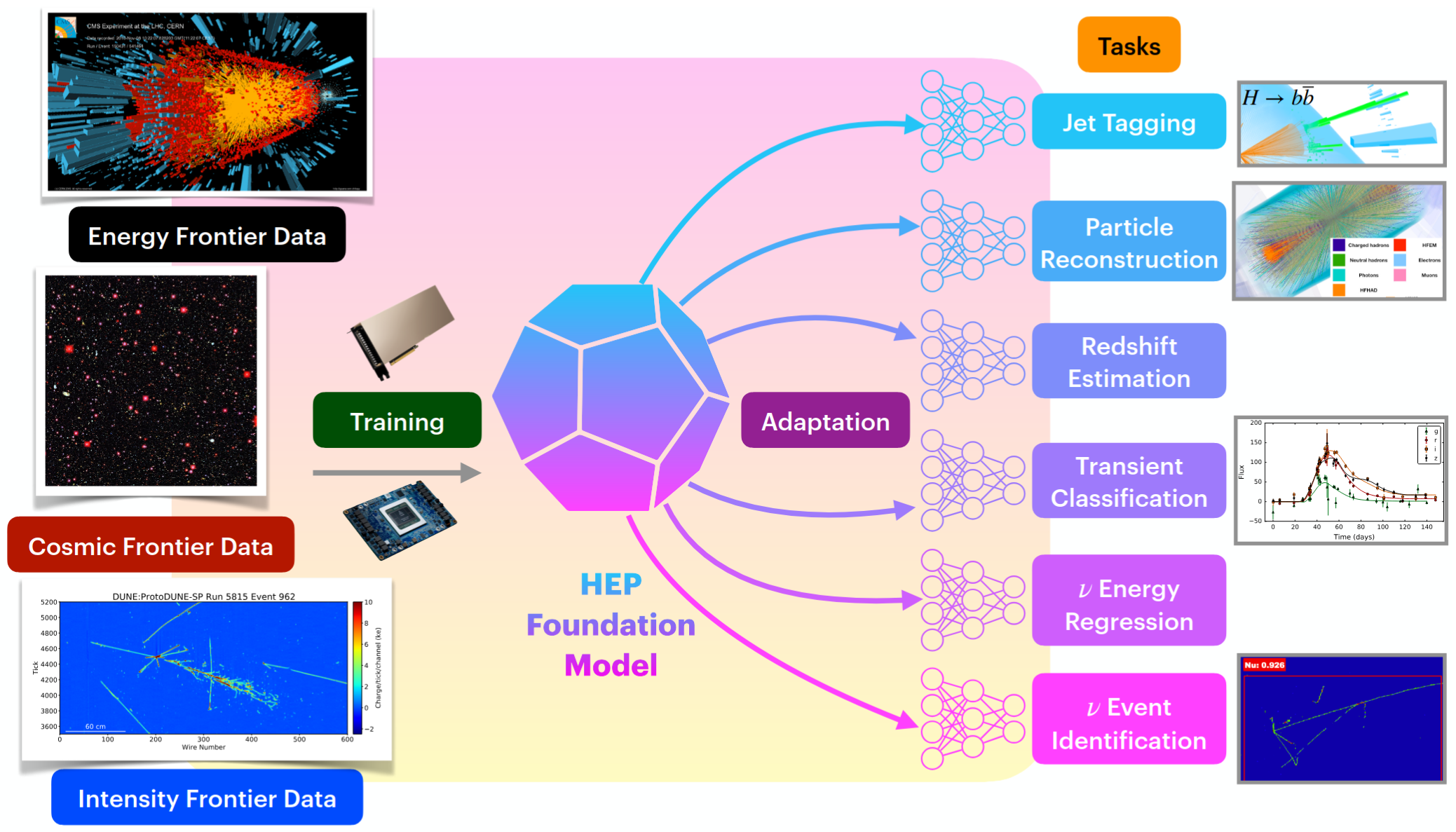}
    \caption{A cross-domain, multimodal foundation model for HEP. }
    \label{fig:HEPFM}
\end{figure}

 To drive progress on these Grand Challenges, several cross-cutting technologies need to be developed or adapted for the particle physics setting. In many cases, progress requires developing new approaches, such as innovative training methods, multi-modality, scaling and adaptability. These AI technologies include:

\paragraph{Foundation Models} are models pretrained in a way that they can be adaptable to a variety of downstream tasks. These models are typically large and trained on massive datasets, necessitating the need to understand \textit{scaling} and the compute requirements for training HEP models. 
The ability to accommodate multiple data modalities (e.g. cross-detector compatibility, accommodating different detector components and data representations, interfacing with human language) is crucial. Further R\&D is needed to find effective methods to fuse information across data modalities to extract general features reusable for high precision, complex particle physics tasks. 
Several such models will be developed, e.g. for adaptable reconstruction, population discrimination, and physics modeling of detector response and particle trajectories that are needed across the Grand Challenges. Inference on such models is also a challenge, especially when aiming to adapt toward low latency systems like triggers, likely necessitating inference as a service systems. Developing the architecture and training procedures for foundation models requires R\&D for physics data types, which extends beyond what has been done in industries, due to the unique data types for high precision physics inference (e.g. data containing millions of events, with millions of sensors per event, and 3D images with billions of voxels). Finally, much of the present research on foundation models for particle physics focuses within each science domain. One large foundation model (Figure~\ref{fig:HEPFM}) that covers all frontiers of particle physics research with a common physics background knowledge will require future R\&D.

 \paragraph{Fast Data Generation / Simulation} to provide rapid, adaptable, and ideally on demand data creation for widespread use in HEP analysis such as training and adapting models, statistical inference, and two-sample testing for data quality. These tools are crucial for design optimization, operations, and analysis (Grand Challenges 1, 3, and 4). Building these tools will require generative AI techniques such as diffusion models or flow-based methods that are adapted to physics data structures, as well as standard simulations that are accelerated through large-scale compute systems and workflow automation. Generative AI methods are necessary for the fast simulation to approximate the full simulation using a limited subset of mathematical operations enabling acceleration on GPUs or other coprocessors. Ensuring the fidelity of these new tools will require robust validation procedures using a large set of curated measurements from HEP experiments.

\paragraph{Differentiable Simulation} corresponds to integration of simulators into automatic differentiation frameworks. By making such codes differentiable, they become amenable to gradient-based optimization and capable of solving a spectrum of inverse problems. Applications concerning Grand Challenges 1, 3 and 4 include sensitivity analysis, automated detector calibration and reconstruction, design optimization, and simulation-based inference. Via chain rules, differentiable simulators can become a greater differentiable pipeline (e.g. extending to reconstruction) and effectively enabling an end-to-end optimization.

 \paragraph{Neural simulation-based inference} includes a variety of methods at the conjunction of simulation, probabilistic modeling, and deep learning and generative AI. These methods enable high-dimensional Bayesian and Frequentist statistical inference that can provide significant improvements in sensitivity compared to prior methods (e.g. histogram-based solutions).

\paragraph{Anomaly detection} applies unsupervised and weakly supervised methods to identify deviations from dominant data distributions without relying on predefined signal hypotheses. It can be used for model-agnostic searches to search for new unanticipated phenomena, to design automated systems to identify unusual or unexpected events in trigger-level filtering, detector monitoring and operations. 

 \paragraph{Active learning and reinforcement learning} provide paradigms to enable iterative and measurement-informed decision making, e.g. for choosing the next hypothesis to test, choosing the next data quality check to perform, or choosing the next step in a design process.  These technologies have only seen limited use in particle physics so far, but will play key roles in the feedback and decision making process envisioned here including design optimization (Grand Challenge 1), autonomous facilities and experiments including accelerators, telescopes, etc (Grand Challenge 3) as well as hyperparameter optimization in training AI models.

 \paragraph{Agentic systems} benefit from the strong reasoning capabilities of modern LLMs, the strong coding abilities of modern LLMs, and the LLM's ability to run external tools and interpret such codes output. Using such tools ``off-the-shelf'' or exploring potential fine-tuning (e.g. via reinforcement learning) of such systems for HEP data can provide the critical outer-loop orchestration of workflows and design planning that can drive more automated decision making across all of the Grand Challenges.

 \paragraph{AI Training and Inference as a Service} is the backbone for any AI-native and -driven experimentation and data analysis. Without ready access and efficient usage of of AI-focused computing resources any progress enabled by AI methods would be hindered. Integrating a new workflow paradigm, i.e. offloading AI compute to an external service or service provider, such as those resources hosted at national laboratories, is essential to reap the highest benefit from AI-native methods. 

\paragraph{On-edge AI} focuses on the development and deployment of AI models on platforms such as FPGAs and ASICs where resources are constrained. These technologies seek to provide inference capabilities for increasingly complex models in systems where the latency and/or resources are highly constrained, such as the trigger at colliders or in systems with remote sensing capabilities. Technology developments focus on areas such as compression to reduce the resource footprint and faster inference, while maintaining physics performance. In addition, developments on platforms located on the detectors often require the devices and algorithms to consider radiation-hardness. 

\paragraph{Digital Twins} are models of physical experiments that are realistic, dynamic and can be queried, adapted, or optimized as if they were the actual experiment. These models are the enabling substrate for AI-native experiments, making it possible to turn the experiment into a learning system. Digital twins are key to enabling agentic AI and reinforcement learning, since they provide tight feedback loops and verifiability.
 
\newpage
%\section{Infrastructure Requirements - TO BE RENAMED Advanced Cyberinfrastructure for AI-Native Experimental HEP(2 pages?)}
%\section{Advanced Cyberinfrastructure for AI-Native Experimental HEP(2 pages?)}
\section{Advanced AI Cyberinfrastructure}
%(2 pages?)
\label{sec:infrastructure}
% [2 pages]

%[Lead: Taylor? for first draft, with Ian, Johannes?, Kaushik?, Lothar, Kazu. Verena will reach out to people.]
%[Giordon - for any inference-as-a-service parts at least, facility-related questions]
%
%[\textcolor{red}{BrianB rewrite of opening}, \textcolor{blue}{Thanks Brian, I made a few additions - Taylor}]

The resources required to support the \textbf{new vision of AI integration} laid
out in these grand challenges will be substantial. They include a massive
increase in access to \textbf{``AI-ready'' data}, as well as support for data
management and curation. They also encompass \textbf{large-scale compute} systems
capable of performing AI training on HEP datasets, serving foundation models for
inference, and hosting LLM for agent-driven experiment workflows including detector
calibration, simulation, reconstruction, and assisted analysis. These end-to-end
science capabilities must be supported with sufficient resources to accelerate
discovery.

Realizing our vision also requires continuous awareness of, and adaptation to, the rapidly evolving \textbf{global cyberinfrastructure} on which data‑intensive science depends. Emerging trends in networking, hardware, and distributed system design must explicitly inform long‑term planning for experiments, software, and operations. Agentic AI architectures will be essential for managing these heterogeneous, globally distributed systems, enabling coordinated optimization of workflows, resources, and policies across institutions while maintaining resilience and scientific integrity.

Particle physics brings \textbf{unique opportunities and skills to the national cyberinfrastructure} (CI). The experiment \textbf{exascale data} is organized, managed, transferred, accessed and accounted for using distributed management systems that are able to integrate scientific and commercial storage systems.
% BrianB commented out this sentence -- it's unclear what strict naming conventions has to do with the vision of AI-ready CI?  Is the reference to "findability" a nod to FAIR?
% Datasets also follow strict naming conventions to ease findability across the full dataset.
In addition, the tradition in HEP of large experiments distributed across many institutions has resulted in \textbf{decades of expertise in distributed computing}; the computing approach has reflected the organizational infrastructure, resulting in valuing portability between sites, common APIs/services, and workloads that can be partitioned effectively across sites.  The HEP community is unparalleled in its ability to be a CI ``omnivore", leveraging almost any conceivable computing infrastructure. These abilities predate, and contributed technology to, commercial cloud computing.  \textbf{HEP's scale in data volume and data flow complexity is unparalleled}: while the large LHC experiments move over an exabyte a year between dozens of computing centers across international boundaries, even smaller experiments leverage the same common services to cross boundaries.  Given the aggregation of researchers' workloads by experiment, modest investments to have experiments adopt new services can impact thousands of scientists and serve as a premier training ground for a new, AI-enabled workforce. 
% BrianB: the use of cloud resources isn't exactly groundbreaking partnerships.  If we want to highlight something different/novel, we should reference the ATLAS work with Google which leveraged unique services (BigTable?) that are not found in federally-funded CI.
% However, my comment is late in the game so I'll leave it untouched for now.
\textbf{Partnerships with industry} are critical to enable \textbf{access to state-of-the-art AI tools and technologies}: the HEP community has extensive experience working  with industry, particularly through subscriptions for resources and services that can be integrated in the HEP cyberinfrastructure.

These historical strengths mean the community (a) can quickly leverage new resources, (b) is skilled at adopting new interfaces, and (c) ensures that modest investments have amplified impact on science. These attributes make the \textbf{HEP community the ideal launchpad for a coordinated national AI infrastructure}.
Building on the HEP community's partnership with diverse funding agencies, we envision leveraging DOE assets like the Leadership Computing Facilities (LCF) and the American Science Cloud (AmSC) at the largest scale and NSF's breadth of infrastructure across the US university ecosystem.
% BrianB: this concept is not really unpacked?
Furthermore, engagement with industry is essential, to create hybrid models that integrate commercial best practices.

Realizing this vision requires \textbf{complementing the current paradigm} of ``distributed batch processing on massive datasets'' \textbf{with new services} that allow massive, bursty training of foundation models at the largest of scales, training of large ensembles of medium-sized models for experimentation with new AI methods (optimal for distributing over all available resources), and the always-on, low-latency inference required by agentic or high-throughput inference workflows.  Combined, these represent a challenge that, if successful, \textbf{leverages and amplifies the value of the existing federal investments} and is a key enabler for the grand challenges.

\paragraph{Data Curation, Governance, and Open Access}
Data is the fuel for foundation models and AI at scale, but reconstructed physics data (interaction events) is rarely ``AI-ready''. To effectively integrate agentic AI, experiments need to curate and preserve all forms of data, ranging from scientific datasets to operational artifacts such as documentation, logs, design records, and publications, in AI-ready formats throughout the lifecycle. Ensuring that this information is structured is key to building a truly AI-native experiment. To enable cross-experiment AI ingestion and empower agents to easily understand and perform data analysis in support of exploratory interactions driven by scientists, the infrastructure must support a dedicated \textbf{Data Curation} layer that transforms these events into coherent, community-defined formats optimized for modern machine learning pipelines.

\begin{itemize}
    \item \textbf{Federated Data Lakehouse and Open Data:} We envision a federated ``Data Lakehouse'' that enables data-level queries to extract specific variables or slices of datasets across distributed facilities. This architecture must support \textbf{Open Data} principles and provide secure access to high-value community datasets while respecting data sovereignty and ownership policies.
    
    \item \textbf{Coherent Representation and Formats:} To enable cross-experiment foundation models, data must be curated into coherent, self-describing representations (e.g., tokenized sequences, vector embeddings) with standardized interfaces, ensuring a model trained on DUNE data can be technically interoperable with analysis tools developed for the LHC.  The approaches to ``tokenize'' event data are rapidly evolving and this area would benefit immensely from US leadership.
    
    \item \textbf{Automated Curation Pipelines:} The infrastructure must provide standardized, containerized workflows to ingest, clean, and annotate interaction events. This includes automated metadata extraction to ensure all data is FAIR (Findable, Accessible, Interoperable, and Reusable) and discoverable by AI agents.
\end{itemize}

Any one of these items is challenging -- however, given the exabyte scale and variety of HEP datasets (a single experiment may have tens of thousands of datasets) and complexity (an event can result in thousands of derived quantities), the opportunity for impact is unprecedented.

\paragraph{Computational Resources and Hardware Adaptation}
The infrastructure must support distinct modalities of AI computation, leverage the investment in existing physics codes to produce new datasets, and use emerging hardware architectures:

\begin{itemize}
    \item \textbf{Training and Fine-Tuning at Scale:} Developing physics foundation models 
    %(Grand Challenge 4~\ref{sec:GC4}) 
    requires access to leadership-class computing facilities (such as NSF's LCCF, expected to begin operations in 2026) capable of massive parallel training. This infrastructure must support flexible execution environments for training complex multi-modal architectures and automated hyper-parameter tuning as a service.

    \item \textbf{Increasing Computational Throughput:} The HEP community has a breadth of models to train, optimize, and use for large-scale processing.  It is uniquely positioned to use all available resources, including mid-scale and university-level investments.
    
    \item \textbf{Adapting Simulation for AI-Centric Hardware:} While traditional HEP simulations heavily rely on double-precision (FP64) arithmetic, modern AI-accelerated hardware increasingly prioritizes mixed- and low-precision capabilities (e.g., FP16, FP8). To maximize scientific throughput on next-generation supercomputers, we must invest in refactoring simulation codes to leverage these lower-precision units, coupled with AI-enhanced or AI-supplemented simulations that do not compromise physical accuracy.
    
    \item \textbf{Agent-Driven Simulation and Orchestration:} Enabling ``Agentic Science'' 
    %(Grand Challenge 3~\ref{sec:GC3}) 
    requires infrastructure that allows AI agents to dynamically provision simulation resources (e.g., event generators, detector simulations) in response to active learning loops. This requires a ``Workflow as a Service'' capability where simulation containers can be spun up on-demand to test hypotheses generated by an AI agent.

    \item \textbf{Long-Running Inference Services:} Deploying AI agents, foundation models, and AI-native simulation and data analysis codes requires persistent, low-latency inference endpoints. The infrastructure must provide stable hosting for open-source foundation models (e.g., Llama, Mistral) while simultaneously offering secure, integrated access to leading proprietary industry models. This enables researchers to query the state-of-the-art—whether open or commercial—instantaneously within their analysis loops.

    \item \textbf{Integration of edge with distributed resources:} Real-time AI processing enables autonomous experiments, but also needs to be complemented by access to distributed data and computational resources at scale for tasks such as AI training and inference, as well as for optimization and validation using simulation or a digital twin. 
\end{itemize}

\paragraph{Leveraging the American Science Cloud (AmSC) for the Genesis Mission}
The developing \textbf{American Science Cloud (AmSC)} offers a unique opportunity to build these capabilities at a national scale. If the High Energy Physics community actively engages with AmSC during its formation, the resulting infrastructure and tools would serve many of the specialized purposes we require, effectively providing a platform for the \textbf{Genesis Mission}.

\begin{itemize}
    \item \textbf{Unified Inference for Open Source and Industry Models:} AmSC plans to provide the ``Inference as a Service'' layer described above, managing the complexity of serving long-running open-source models while handling the authentication and subscriptions required for industry model access.
    
    \item \textbf{Genesis Mission Alignment:} By providing a unified platform for AI model development and agentic workflows, AmSC directly supports the Genesis Mission's goal of accelerating scientific discovery through AI integration. 
    %It provides the necessary ``Information Space Laboratory'' described in Section 7, enabling rapid iteration on scientific ideas.
    
    \item \textbf{Agentic Framework and MCP:} AmSC is designing an \textbf{Intelligent Interfaces} layer to support AI agents via \textbf{Model Context Protocol (MCP)} servers. This allows an AI agent to ``call'' a supercomputer simulation tool as easily as a Python function, enabling the fully automated ``self-driving'' cycles envisioned for future colliders.
    
    \item \textbf{Collaboration on Use Cases:} To ensure this infrastructure meets the unique needs of particle physics, the HEP community must engage in co-designing specific use cases with AmSC and industry partners. This collaboration will drive the development of features such as low-latency triggers and petabyte-scale data loaders.
\end{itemize}

% Meeting Notes:

% Inference
% AmSC plans to provide access to long-running LLMs that are both open source and industry
% Access to data and tools - open data

% why it would be useful for Genesis?
% Might also want to include different areas?

% Connection to industry/vendors - or some references to possibilities
% - mixed, low- precision cores and algorithmic improvements on their hardware.
% Hardware, AI tools and models
% Subscriptions for all sorts of resources and services

% Collaboration on use cases 

% Data curation to produce datasets that can be used to produce models
% Data access, ownership, etc
% Coherent format, data representation, data interface, etc

\newpage
\section{Conclusion}
\label{sec:conclusion}
This whitepaper reflects the collective input, expertise, and ambition of a broad 
cross-section of the U.S. particle physics community. Contributors spanning many 
experiments, frontiers, and technical domains have articulated a compelling 
vision for how AI can transform experimental particle physics: accelerating 
discovery, enabling new scientific capabilities across the full experimental 
lifecycle, and empowering a new generation of scientists with the skills needed 
to lead in an AI-enabled research landscape.

At the same time, this vision is inherently dynamic. AI technologies, 
experimental programs, and scientific goals will continue to evolve, as will our 
understanding of where AI delivers the greatest impact. Progress on the projects 
and ideas outlined here will inform new directions, reveal unanticipated 
opportunities, and refine existing approaches. Sustaining scientific impact will, 
therefore, require an iterative strategy that evolves alongside advances in both 
AI and experimental particle physics. 
Furthermore, as the community continues to explore the overall AI landscape, we 
recognize that several broader topics raised in community feedback--such as
sustainability, responsible use of emerging AI technologies, international
engagement, and outreach--extend beyond the present technical scope of this
document. These areas merit thoughtful consideration in future community
efforts, where they can be addressed with the depth and perspective they
require. 

For all of the above reasons, this whitepaper is best viewed as the foundation of a living 
community vision, rather than a static document. Periodic updates would allow the community to incorporate new insights, assess 
progress, and re-calibrate priorities. Through a sustained, national-scale 
collaboration that unites laboratories, universities, and shared AI 
infrastructure around these Grand Challenges, the community can translate this 
evolving vision into coordinated action and lasting scientific impact.

\paragraph{Acknowledgements:} We gratefully acknowledge support from the Simons
Foundation and from the National Science Foundation under cooperative agreement
PHY-2323298 (IRIS-HEP), which enabled the organizational process to assemble
this community vision on an accelerated timeline. The National Science
Foundation and the Department of Energy also provided support for individual
contributors participating in this effort.

\newpage 
\bibliographystyle{unsrt}
\bibliography{aiml2026}

@misc{iris-hep,
  title = "{Website - Institute for Research and Innovation in Software for High Energy Physics (IRIS-HEP)}",
  howpublished = "\url{http://iris-hep.org}"
}

@misc{iris-hep-nsf-award,
  title = "{National Science Foundation Cooperative Agreement OAC-1836650}",
  howpublished = "\url{https://www.nsf.gov/awardsearch/showAward?AWD_ID=1836650&HistoricalAwards=false}"
}

@misc{iris-hep-nsf-renewal,
  title = "{National Science Foundation Cooperative Agreement PHY-2323298}",
  howpublished = "\url{https://www.nsf.gov/awardsearch/showAward?AWD_ID=2323298&HistoricalAwards=false}"
}

@book{hep_p5_2023,
   title={Exploring the Quantum Universe: Pathways to Innovation and Discovery in Particle Physics},
   url={http://dx.doi.org/10.2172/2368847},
   DOI={10.2172/2368847},
   institution={Office of Scientific and Technical Information (OSTI)},
   author={Murayama, Hitoshi and Asai, Shoji and Heeger, Karsten and Ballarino, Amalia and Bose, Tulika and Cranmer, Kyle and Cyr-Racine, Francis-Yan and Demers, Sarah and Geddes, Cameron and Gershtein, Yuri and Heinemann, Beate and Hewett, JoAnne and Huber, Patrick and Mahn, Kendall and Mandelbaum, Rachel and Maricic, Jelena and Merkel, Petra and Monahan, Christopher and Onyisi, Peter and Palmer, Mark and Raubenheimer, Tor and Sanchez, Mayly and Schnee, Richard and Seidel, Sally and Seo, Seon-Hee and Thaler, Jesse and Touramanis, Christos and Vieregg, Abigail and Weinstein, Amanda and Winslow, Lindley and Yu, Tien-Tien and Zwaska, Robert},
   year={2023},
   month=jun }

@book{NASEM2025HiggsBeyond28839,
  author    = {{National Academies of Sciences, Engineering, and Medicine}},
  title     = {Elementary Particle Physics: The Higgs and Beyond},
  year      = {2025},
  publisher = {National Academies Press},
  address   = {Washington, DC},
  doi       = {10.17226/28839},
  url       = {https://doi.org/10.17226/28839}
}

@misc{a3d3-web,
  title = {{Website - A3D3: Accelerated Artificial Intelligence Algorithms for Data-Driven Discovery}},
  howpublished = "\url{https://a3d3.ai}"
}

@misc{a3d3-postbac,
  title = {{Website - A3D3: Accelerated Artificial Intelligence Algorithms for Data-Driven Discovery - Postbac Program}},
  howpublished = "\url{https://a3d3.ai/education-and-outreach/postbac/}"
}

@misc{a3d3-nsf-award,
  title = "{National Science Foundation Cooperative Agreement OAC-2117997}",
  howpublished = "\url{https://www.nsf.gov/awardsearch/show-award?AWD_ID=2117997}"
}

@misc{watchep,
  title = {{Website - Western Advanced Training for Computational High-Energy Physics (WATCHEP)}},
  howpublished = "\url{https://watchep.org}"
}

@misc{tac-hep,
  title = {{Website - Training to Advance Computational High Energy Physics in the Exascale Era (TAC-HEP)}},
  howpublished = "\url{https://tac-hep.org}"
}

@misc{hep-cce,
  title = {{Website - High Energy Physics - Center for Computational Excellence (HEP-CCE))}},
  howpublished = "\url{https://www.anl.gov/hep-cce}"
}

@misc{c2-the-p2,
  title = {{Website - Chicagoland Computational Traineeship in High Energy Particle Physics ($C^2$ the $P^2$}},
  howpublished = "\url{https://www.c2thep2.org}"
}

@misc{DOE-RFI-workforce-development,
  title = {{DE-SC-26-016: Request for Information (RFI) on Mobilizing Talent for the Genesis Mission and Developing an American Workforce to Advance Artificial Intelligence (AI) for Science and Engineering}},
  howpublished = "https://sam.gov/workspace/contract/opp/1f6ee2898b724568b4816de787d0b8f6/view"
}

@misc{quarknet,
  title = {{Website - QuarkNet}},
  howpublished = "\url{https://quarknet.org}"
}

@misc{ml4fp-2025,
  title = {{Website - Machine Learning for Fundamental Physics School (ML4FP) 2025 }},
  howpublished = "\url{https://indico.physics.lbl.gov/event/3174/}"
}

@misc{moller-jlab,
      title={The MOLLER Experiment: An Ultra-Precise Measurement of the Weak Mixing Angle Using M{\o}ller Scattering}, 
      author={MOLLER Collaboration and J. Benesch and P. Brindza and R. D. Carlini and J-P. Chen and E. Chudakov and S. Covrig and M. M. Dalton and A. Deur and D. Gaskell and A. Gavalya and J. Gomez and D. W. Higinbotham and C. Keppel and D. Meekins and R. Michaels and B. Moffit and Y. Roblin and R. Suleiman and R. Wines and B. Wojtsekhowski and G. Cates and D. Crabb and D. Day and K. Gnanvo and D. Keller and N. Liyanage and V. V. Nelyubin and H. Nguyen and B. Norum and K. Paschke and V. Sulkosky and J. Zhang and X. Zheng and J. Birchall and P. Blunden and M. T. W. Gericke and W. R. Falk and L. Lee and J. Mammei and S. A. Page and W. T. H. van Oers and K. Dehmelt and A. Deshpande and N. Feege and T. K. Hemmick and K. S. Kumar and T. Kutz and R. Miskimen and M. J. Ramsey-Musolf and S. Riordan and N. Hirlinger Saylor and J. Bessuille and E. Ihloff and J. Kelsey and S. Kowalski and R. Silwal and G. De Cataldo and R. De Leo and D. Di Bari and L. Lagamba and E. NappiV. Bellini and F. Mammoliti and F. Noto and M. L. Sperduto and C. M. Sutera and P. Cole and T. A. Forest and M. Khandekar and D. McNulty and K. Aulenbacher and S. Baunack and F. Maas and V. Tioukine and R. Gilman and K. Myers and R. Ransome and A. Tadepalli and R. Beniniwattha and R. Holmes and P. Souder and D. S. Armstrong and T. D. Averett and W. Deconinck and W. Duvall and A. Lee and M. L. Pitt and J. A. Dunne and D. Dutta and L. El Fassi and F. De Persio and F. Meddi and G. M. Urciuoli and E. Cisbani and C. Fanelli and F. Garibaldi and K. Johnston and N. Simicevic and S. Wells and P. M. King and J. Roche and J. Arrington and P. E. Reimer and G. Franklin and B. Quinn and A. Ahmidouch and S. Danagoulian and O. Glamazdin and R. Pomatsalyuk and R. Mammei and J. W. Martin and T. Holmstrom and J. Erler and Yu. G. Kolomensky and J. Napolitano and K. A. Aniol and W. D. Ramsay and E. Korkmaz and D. T. Spayde and F. Benmokhtar and A. Del Dotto and R. Perrino and S. Barkanova and A. Aleksejevs and J. Singh},
      year={2014},
      eprint={1411.4088},
      archivePrefix={arXiv},
      primaryClass={nucl-ex},
      url={https://arxiv.org/abs/1411.4088}, 
}

@misc{solid-jlab,
      title={The Solenoidal Large Intensity Device (SoLID) for JLab 12 GeV}, 
      author={John Arrington and Jay Benesch and Alexandre Camsonne and Jimmy Caylor and Jian-Ping Chen and Silviu Covrig Dusa and Alexander Emmert and George Evans and Haiyan Gao and J. Ole Hansen and Garth M. Huber and Sylvester Joosten and Vladimir Khachatryan and Nilanga Liyanage and Zein-Eddine Meziani and Michael Nycz and Chao Peng and Michael Paolone and Whit Seay and Paul A. Souder and Nikos Sparveris and Hubert Spiesberger and Ye Tian and Eric Voutier and Junqi Xie and Weizhi Xiong and Zhenyu Ye and Zhihong Ye and Jixie Zhang and Zhiwen Zhao and Xiaochao Zheng},
      year={2023},
      eprint={2209.13357},
      archivePrefix={arXiv},
      primaryClass={nucl-ex},
      url={https://arxiv.org/abs/2209.13357}, 
}

@misc{codas-hep,
  title = {{Website - Computational and Data Science Training for High Energy Physics (CoDaS-HEP)}},
  howpublished = "\url{https://codas-hep.org}"
}

@misc{iaifi,
  title = {{Website - NSF AI Institute for Artificial Intelligence and Fundamental Interactions (IAIFI)}},
  howpublished = "\url{https://iaifi.org}"
}

@misc{iaifi-summer-school,
  title = {{Website - IAIFI Summer School}},
  howpublished = "\url{https://iaifi.org/phd-summer-school.html}"
}

@misc{sbi-website,
  title = {{Website - Simulation Based Inference}},
  howpublished = "\url{https://simulation-based-inference.org}"
}

@misc{shanahan2022snowmass2021computationalfrontier,
      title={{Snowmass 2021 Computational Frontier CompF03 Topical Group Report: Machine Learning}}, 
      author={Phiala Shanahan and Kazuhiro Terao and Daniel Whiteson},
      year={2022},
      eprint={2209.07559},
      archivePrefix={arXiv},
      primaryClass={physics.comp-ph},
      url={https://arxiv.org/abs/2209.07559}, 
}

@misc{hepml-living-review-website,
  title = {{Website - A Living Review of Machine Learning for Particle Physics}},
  howpublished = "\url{https://iml-wg.github.io/HEPML-LivingReview/}"
}

@misc{feickert2021livingreviewmachinelearning,
      title={A Living Review of Machine Learning for Particle Physics}, 
      author={Matthew Feickert and Benjamin Nachman},
      year={2021},
      eprint={2102.02770},
      archivePrefix={arXiv},
      primaryClass={hep-ph},
      url={https://arxiv.org/abs/2102.02770}, 
}

@misc{dune-tech,
  title = {{Website - DUNE Training ExperienCe Hub (DUNE-TECH)}},
  howpublished = "\url{https://dune-tech.rice.edu}"
}

@misc{kaganheinrichtalk,
    author = {Kagan, M. and Heinrich, L.},
    title = "{Presentation - New Frontiers in AI for Fundamental Physics}",
    howpublished = "{Based on presentation \url{https://indico.cern.ch/event/1604420/contributions/6760798/}}"
}

@article{lsst-desc,
  doi = {10.5281/ZENODO.18319953},
  
  url = {https://zenodo.org/doi/10.5281/zenodo.18319953},

author = {{LSST Dark Energy Science Collaboration} and Aubourg, Eric and Avestruz, Camille and Becker, Matthew R. and Biswas, Biswajit and Biswas, Rahul and Bolliet, Boris and Bolton, Adam S. and Bom, Clecio R. and Bonnet-Guerrini, Raphaël and Boucaud, Alexandre and Campagne, Jean-Eric and Chang, Chihway and Ćiprijanović, Aleksandra and Cohen-Tanugi, Johann and Coughlin, Michael W. and Crenshaw, John Franklin and Cuevas-Tello, Juan C. and de Vicente, Juan and Digel, Seth W. and Dillmann, Steven and Romero, Mariano Javier de León Dominguez and Drlica-Wagner, Alex and Erickson, Sydney and Gagliano, Alexander T. and Georgiou, Christos and Ghosh, Aritra and Grayling, Matthew and Grishin, Kirill A. and Heavens, Alan and House, Lindsay R. and Ishak, Mustapha and Kabalan, Wassim and Kannawadi, Arun and Lanusse, François and Leonard, C. Danielle and Léget, Pierre-François and Lochner, Michelle and Mao, Yao-Yuan and Melchior, Peter and Merz, Grant and Millon, Martin and Möller, Anais and Narayan, Gautham and Omori, Yuuki and Peiris, Hiranya and Perreault-Levasseur, Laurence and Malagón, Andrés A. Plazas and Ramachandra, Nesar and Remy, Benjamin and Roucelle, Cécile and Ruiz-Zapatero, Jaime and Schuldt, Stefan and Sevilla-Noarbe, Ignacio and Shah, Ved G. and Starkenburg, Tjitske and Thorp, Stephen and Cipriano, Laura Toribio San and Tröster, Tilman and Trotta, Roberto and Venkatraman, Padma and Wasserman, Amanda and White, Tim and Zeghal, Justine and Zhang, Tianqing and Zhang, Yuanyuan},
  
  title = {Opportunities in AI/ML for the Rubin LSST Dark Energy Science Collaboration},
  
  publisher = {Zenodo},
  
  year = {2026},
  
  copyright = {Creative Commons Attribution 4.0 International}
}

@techreport{osti_2280968,
  author       = {US Department of Energy (USDOE)},
  title        = {A New Era of Discovery: The 2023 Long Range Plan for Nuclear Science},
  institution  = {US Department of Energy (USDOE), Washington, DC (United States). Office of Science},
  doi          = {10.2172/2280968},
  url          = {https://www.osti.gov/biblio/2280968},
  place        = {United States},
  year         = {2023},
  month        = {10}}

@misc{nsf-mps-2025,
      title={The Future of Artificial Intelligence and the Mathematical and Physical Sciences (AI+MPS)}, 
      author={Andrew Ferguson and Marisa LaFleur and Lars Ruthotto and Jesse Thaler and Yuan-Sen Ting and Pratyush Tiwary and Soledad Villar and E. Paulo Alves and Jeremy Avigad and Simon Billinge and Camille Bilodeau and Keith Brown and Emmanuel Candes and Arghya Chattopadhyay and Bingqing Cheng and Jonathan Clausen and Connor Coley and Andrew Connolly and Fred Daum and Sijia Dong and Chrisy Xiyu Du and Cora Dvorkin and Cristiano Fanelli and Eric B. Ford and Luis Manuel Frutos and Nicolás García Trillos and Cecilia Garraffo and Robert Ghrist and Rafael Gomez-Bombarelli and Gianluca Guadagni and Sreelekha Guggilam and Sergei Gukov and Juan B. Gutiérrez and Salman Habib and Johannes Hachmann and Boris Hanin and Philip Harris and Murray Holland and Elizabeth Holm and Hsin-Yuan Huang and Shih-Chieh Hsu and Nick Jackson and Olexandr Isayev and Heng Ji and Aggelos Katsaggelos and Jeremy Kepner and Yannis Kevrekidis and Michelle Kuchera and J. Nathan Kutz and Branislava Lalic and Ann Lee and Matt LeBlanc and Josiah Lim and Rebecca Lindsey and Yongmin Liu and Peter Y. Lu and Sudhir Malik and Vuk Mandic and Vidya Manian and Emeka P. Mazi and Pankaj Mehta and Peter Melchior and Brice Ménard and Jennifer Ngadiuba and Stella Offner and Elsa Olivetti and Shyue Ping Ong and Christopher Rackauckas and Philippe Rigollet and Chad Risko and Philip Romero and Grant Rotskoff and Brett Savoie and Uros Seljak and David Shih and Gary Shiu and Dima Shlyakhtenko and Eva Silverstein and Taylor Sparks and Thomas Strohmer and Christopher Stubbs and Stephen Thomas and Suriyanarayanan Vaikuntanathan and Rene Vidal and Francisco Villaescusa-Navarro and Gregory Voth and Benjamin Wandelt and Rachel Ward and Melanie Weber and Risa Wechsler and Stephen Whitelam and Olaf Wiest and Mike Williams and Zhuoran Yang and Yaroslava G. Yingling and Bin Yu and Shuwen Yue and Ann Zabludoff and Huimin Zhao and Tong Zhang},
      year={2025},
      eprint={2509.02661},
      archivePrefix={arXiv},
      primaryClass={cs.AI},
      url={https://arxiv.org/abs/2509.02661}, 
}

@misc{ai2ased,
  author      = {Daniel Anglés-Alcázar and Animashree Anandkumar and Joshua C Agar and Bedrich Benes and Ying Ding and Wei Ding and Lili Du and Jennifer Dy and Baskar Ganapathysubramanian and Krishna Garikipati and Jing Gao and Omar Ghattas and Jane Greenberg and Paul C Hanson and Marti Hearst and Phil Harris and Shirley Ho and Mingyi Hong and Shih-Chieh Hsu and Shuiwang Ji and Anuj Karpatne and Carl Kingsford and Vipin Kumar and L. Ruby Leung and Edgar Lobaton and Madhav V. Marathe and Nirav Merchant and Peetak Mitra and Mark S. Neubauer and Xia Ning and Yuhan Douglas Rao and Balaji Rajagopalan and Xinghua Shi and Jianhui Sun and Brandon Sutherland and Eric Toberer and Wei Wang and Jianwu Wang and Aidong Zhang},
  title       = {{AI to Accelerate Science and Engineering Discovery}},
  institution = {National Science Foundation},
  year        = {2023},
  type        = {},
  number      = {},
  url         = {https://indico.cern.ch/event/1325897/}
}

@misc{iaifi2025,
  title = {{A Virtuous Cycle: Generative AI and Discovery in the Physical Sciences}},
  howpublished = "\url{https://mit-genai.pubpub.org/pub/ewp5ckmf/release/2}"
}

@misc{sage,
  title = {{Website - Seismological Facility for the Advancement of Geoscience (SAGE)}},
  howpublished = "\url{https://www.iris.edu/hq/sage}"
}

@misc{gage,
  title = {{Website - Geodetic Facility for the Advancement of Geoscience (GAGE)}},
  howpublished = "\url{https://www.unavco.org/what-we-do/gage-facility/}"
}

@misc{neon,
  title = {{Website - National Ecological Observatory Network (NEON)}},
  howpublished = "\url{https://www.neonscience.org}"
}

@techreport{NSF25-349,
  author      = {{National Center for Science and Engineering Statistics}},
  title       = {Research Doctorate Recipients' Sources of Financial Support, by Broad Field of Doctorate and Sex: 2024},
  institution = {National Science Foundation},
  year        = {2025},
  type        = {Data Table},
  number      = {NSF 25-349},
  url         = {https://ncses.nsf.gov/pubs/nsf25349/assets/data-tables/tables/nsf25349-tab004-002.pdf}
}

@techreport{ncses2021doctorate,
  author      = {{National Center for Science and Engineering Statistics (NCSES)}},
  title       = {Doctorate Recipients from U.S. Universities: 2020},
  institution = {National Science Foundation},
  year        = {2021},
  type        = {Report},
  number      = {NSF 22-300},
  address     = {Alexandria, VA},
  url         = {https://ncses.nsf.gov/pubs/nsf22300}
}

@misc{pioneercollaboration2025europeanstrategyparticlephysics,
      title={European Strategy for Particle Physics Update -- PIONEER: a next generation rare pion decay experiment}, 
      author={PIONEER Collaboration },
      year={2025},
      eprint={2504.06375},
      archivePrefix={arXiv},
      primaryClass={hep-ex},
      url={https://arxiv.org/abs/2504.06375}, 
}

@misc{beamroadmap,
title={{Accelerator and Beam Physics Roadmap}},
author={Gerald Blazey and others},
year=2023,
type={report},
institution={Department of Energy Office of High Energy Physics},
howpublished="\url{https://science.osti.gov/hep/-/media/hep/pdf/2022/ABP_Roadmap_2023_final.pdf}"
}

@techreport{HEP_ABP_Roadmap_2023,
  title        = {Accelerator and Beam Physics Roadmap 2023},
  institution  = {U.S. Department of Energy, Office of Science, High Energy Physics},
  year         = {2023},
  url          = {https://science.osti.gov/hep/-/media/hep/pdf/2022/ABP_Roadmap_2023_final.pdf},
  note         = {Accessed: 2026-02-19}
}

\end{document}